\documentclass[preprint]{aastex}
\usepackage{longtable}
\usepackage{graphics}
\usepackage{rotating}
\usepackage{amsmath}
\usepackage{natbib}
\usepackage{hyperref}
\def\gtorder{\mathrel{\raise.3ex\hbox{$>$}\mkern-14mu
             \lower0.6ex\hbox{$\sim$}}}
\def\ltorder{\mathrel{\raise.3ex\hbox{$<$}\mkern-14mu
             \lower0.6ex\hbox{$\sim$}}}

\slugcomment{\today}

\shorttitle{Supernova PTF\,11qcj}

\shortauthors{Corsi et al.}

\begin{document}

\title{A multi-wavelength investigation of the radio-loud supernova PTF11qcj and its circumstellar environment}
\author{A.~Corsi\altaffilmark{1,2},
  E.~O.~Ofek\altaffilmark{3},
  A.~Gal-Yam\altaffilmark{3},
  D.~A.~Frail\altaffilmark{4},
  S.~R. Kulkarni\altaffilmark{5},
  D.~B.~Fox\altaffilmark{6},
  M.~M.~Kasliwal\altaffilmark{7},
  M.~Sullivan\altaffilmark{8},
  A.~Horesh\altaffilmark{5},
  J.~Carpenter\altaffilmark{5},
  K.~Maguire\altaffilmark{8},
  I.~Arcavi\altaffilmark{5},
  S.~B.~Cenko\altaffilmark{9,10},
  Y.~Cao\altaffilmark{5},
  K.~Mooley\altaffilmark{5},
  Y.-C.~Pan\altaffilmark{8},
  B.~Sesar\altaffilmark{5},
  A.~Sternberg\altaffilmark{11},
  D.~Xu\altaffilmark{3,12},
  D.~Bersier\altaffilmark{13},
  P.~James\altaffilmark{13},
  J.~S.~Bloom\altaffilmark{10},
  P.~E.~Nugent\altaffilmark{10,14}}
 \altaffiltext{1}{Department of Physics, The George Washington University, 725 21st St, NW, Washington, DC 20052, USA; e-mail: corsi@gwu.edu}
\altaffiltext{2}{LIGO Laboratory, California Institute of Technology, Pasadena, CA 91125, USA}
\altaffiltext{3}{Benoziyo Center for Astrophysics, Weizmann Institute of Science, 76100 Rehovot, Israel}
\altaffiltext{4}{National Radio Astronomy Observatory, P.O. Box O, Socorro, NM 87801}
\altaffiltext{5}{Division of Physics, Mathematics, and Astronomy, California Institute of Technology, Pasadena, CA 91125, USA}
\altaffiltext{6}{Department of Astronomy and Astrophysics, Pennsylvania State University, University Park, PA 16802, USA}
\altaffiltext{7}{Observatories of the Carnegie Institution for Science, 813 Santa Barbara Street, Pasadena, CA 91101, USA}
\altaffiltext{8}{Department of Physics, University of Oxford, Denys Wilkinson Building, Keble Road, Oxford OX1 3RH, UK}
\altaffiltext{9}{NASA Goddard Space Flight Center, Code 685, Greenbelt, MD 20771.}
\altaffiltext{10}{Department of Astronomy, University of California, Berkeley, 601 Campbell Hall, Berkeley, CA 94720}
\altaffiltext{11}{Max-Planck-Institut fur Astrophysik, 85741 Garching, Germany.}
\altaffiltext{12}{Dark Cosmology Centre, Niels Bohr Institute, University of Copenhagen, Juliane Maries Vej 30, 2100 K$\oslash$benhavn $\oslash$, Denmark.}
\altaffiltext{13}{Astrophysics Research Institute, Liverpool John Moores University, Liverpool, UK}
\altaffiltext{14}{Lawrence Berkeley National Laboratory, MS 50B-4206, 1 Cyclotron Road, Berkeley, CA, 94720-8139}

\begin{abstract}
We present the discovery, classification, and extensive panchromatic (from radio to X-ray) follow-up observations of PTF11qcj, a supernova discovered by the Palomar Transient Factory. Our observations with the Karl G. Jansky Very Large Array show that this event is radio-loud: PTF11qcj reached a radio peak luminosity comparable to that of the famous gamma-ray-burst-associated supernova 1998bw ($L_{\rm 5\,GHz}\approx 10^{29}$\,erg\,s$^{-1}$\,Hz$^{-1}$). PTF11qcj is also detected in X-rays with the \textit{Chandra} observatory, and in the infrared band with \textit{Spitzer}. Our multi-wavelength analysis probes the supernova interaction with circumstellar material. The radio observations suggest a progenitor mass-loss rate of $\sim 10^{-4}$\,M$_{\odot}$\,yr$^{-1}\times ({\rm v}_w/1000$\,km\,s$^{-1}$), and a velocity of $\approx 0.3-0.5\,c$ for the fastest moving ejecta (at $\approx 10$\,d after explosion). However, these estimates are derived assuming the simplest model of supernova ejecta interacting with a smooth circumstellar wind, and do not account for possible inhomogeneities in the medium and asphericity of the explosion. The radio data show deviations from such a simple model, as well as a late-time re-brightening. The X-ray flux from PTF11qcj is compatible with the high-frequency extrapolation of the radio synchrotron emission (within the large uncertainties). A light echo from pre-existing dust is in agreement with our infrared data. Our pre-explosion data from the Palomar Transient Factory suggest that a precursor eruption of absolute magnitude $M_r \approx -13$\,mag may have occurred $\approx 2.5$\,yr prior to the supernova explosion. Overall, PTF11qcj fits the expectations from the explosion of a  Wolf-Rayet star. Precursor eruptions may be a feature characterizing the final pre-explosion evolution of such stars.
\end{abstract}

\section{Introduction}
\label{Introduction}
Supernovae (SNe) of type Ib/c are believed to mark the death of massive stars that are stripped of their hydrogen (H), and possibly helium (He) envelopes before explosion \citep[e.g.,][]{Filippenko1997}. Since the discovery of an association between the low-luminosity, long-duration $\gamma$-ray burst (GRB) 980425 and SN\,1998bw \citep{Galama1998,Kulkarni1998,Pian1999}, it has become evident that long duration GRBs are a rare sub-class of type Ib/c SNe \citep[for a recent review, see ][and references therein]{Bloom2006}. Their ejecta are highly energetic and collimated relativistic outflows, emitting in $\gamma$-rays, and likely powered by a central engine (an accreting black hole or a neutron star).

SN\,1998bw showed that events with properties intermediate to ``classic'' high-energy, highly relativistic GRBs, and ordinary non-relativistic Ib/c SNe, do exist and could probably allow us to solve the mystery of the GRB-SN connection if more of them were to be discovered.

After GRB\,980425/SN\,1998bw, a few other GRBs have been reliably associated with nearby SNe \citep{Bloom2006}, that are all of type Ic ``broad-line'' (BL), i.e. showing broad spectral features indicative of high photospheric velocities. After hundreds of Ib/c SNe have been monitored in radio in search for SN\,1998bw-like events, SN\,2009bb has marked the discovery of a Ic SN characterized by a relativistic (radio and X-ray emitting) ejecta, with no associated $\gamma$-rays \citep{Soderberg2010}. 

Radio observations are a key to discover events just on the dividing line between GRBs and ordinary Ib/c SNe \citep{Berger2003,Soderberg2006}. In fact, while the SN optical emission traces the slower explosion debris ($v\approx 0.03-0.1$\,c), synchrotron emission from the fastest ejecta peaks in the radio band (on typical timescales of 10-30\,d since explosion). Radio (and X-ray) emission, because it originates from the interaction between the high velocity SN ejecta and the low-velocity wind from the SN progenitor, can probe the density of the circumstellar medium (CSM), and constrain the wind properties of the SN progenitor. Since massive stars are speculated to undergo strong episodic mass-loss events, radio studies of type Ib/c SNe can help us understand the nature of their progenitors.

Here, we report on the discovery of a radio-loud SN, PTF11qcj, and on the results of our extensive, multi-wavelength follow-up campaign extending till $\approx$1.5\,yr after optical detection. While further observations are on-going, the data gathered during the first year of our campaign allow us to set important constraints on the nature of this event. Our paper is organized as follows. In Section \ref{Observations} we describe the optical discovery, spectral classification, and IR-to-X-ray follow-up observations of this event; in Section \ref{Analysis} we describe our multi-frequency analysis; finally, in Section \ref{conclusion} we present our summary. 
\section{Panchromatic observations}
\label{Observations}
\subsection{SN optical photometry}
\label{Opticalphotometry}
\begin{figure*}
\begin{center}
\includegraphics[width=16.cm]{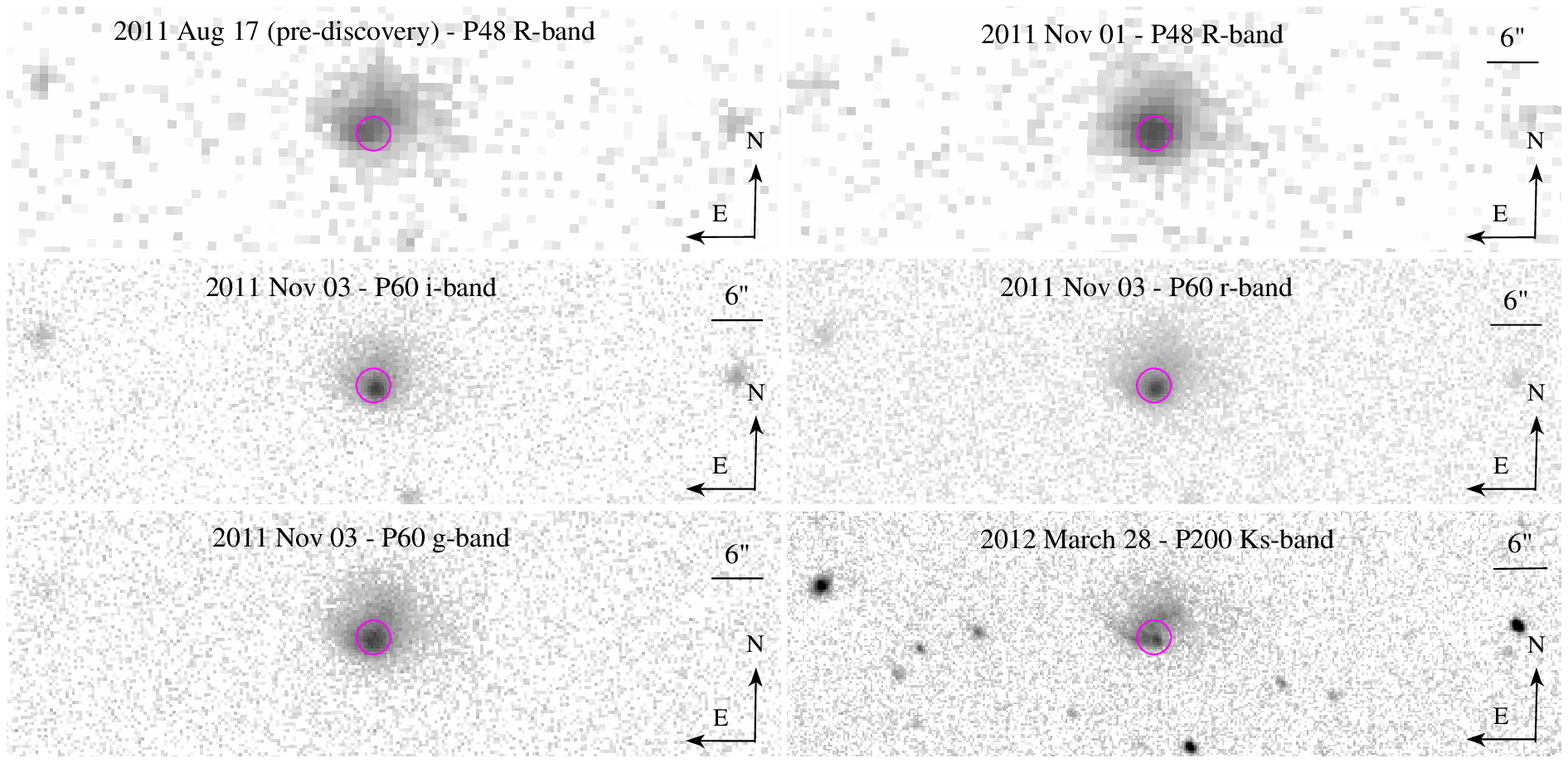}
\caption{P48 Mould-R pre-discovery (upper-left) and discovery (upper-right) images of the PTF11qcj field. We also show multi-color images taken with P60 and P200 during our multi-wavelength follow-up campaign. In all the images, the P48 discovery position of PTF11qcj is marked with a circle of $2$\,\,\arcsec radius. \label{scoperta}}
\end{center}
\end{figure*}

On 2011 November 01 (hereafter all times are given in UTC unless otherwise stated), we discovered PTF11qcj in an $R$-band image from the 48-inch Samuel Oschin telescope at Palomar Observatory (P48), which is routinely used by the Palomar Transient Factory \citep[PTF\footnote{\url{http://ptf.caltech.edu/iptf/}};][]{Law2009,Rau2009}. The SN was also detected by our autonomous discovery and classification framework called {\it Oarical} \citep{Bloom2012}. PTF11qcj is located at $\alpha = 13^{h}13^{m}41.51^{s}$, $\delta = +47^{\circ} 17\arcmin 57\farcs0$ (J2000), at an angular distance of $\approx 0.68\arcsec$ from the nucleus of the galaxy SDSS\footnote{Sloan Digital Sky Survey \citep{York2000}} J131341.57+471757.2, at $z=0.0287$  (Figure \ref{scoperta}). Assuming $H_0=71$\,km\,s$^{-1}$\,Mpc$^{-1}$, $\Omega_M=0.27$, $\Omega_{\Lambda}=0.73$, this redshift correspond to a luminosity distance of $d_L\cong 124$\,Mpc. The SN was visible at a magnitude of $g\approx 17.6$ in a previous exposure taken in $g$-band on 2011 October 23. It was not detected in an $R$-band image taken with the P48 on 2011 August 17.  

\begin{figure}
\begin{center}
\includegraphics[width=8.5cm]{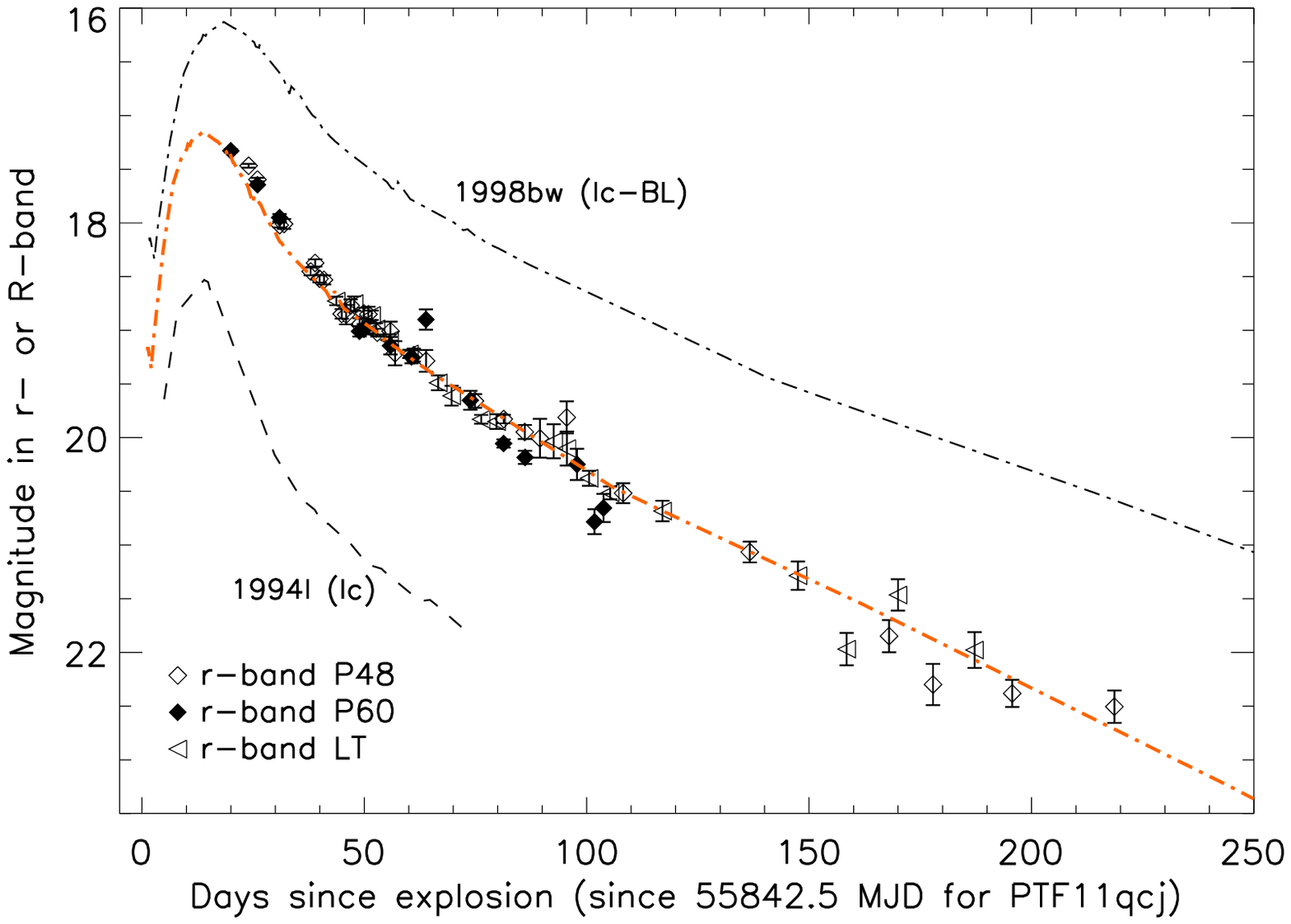}
\caption{The PTF11qcj $r$-band light curve (corrected for Galactic extinction and local host-galaxy extinction) is compared with the light curve of the Ic-BL SN\,1998bw \citep{Clocchiatti2011}, and of the normal Ic SN\,1994I \citep{Richmond1996}. The orange-dashed line is the optical light curve of SN\,1998bw compressed in time by a factor of $1.33$ and scaled in peak magnitude so as to match our P48 observations of PTF11qcj. (See the electronic version of this paper for colors.) \label{lc_mag}}
\end{center}
\end{figure}

Subsequent observations with the P48 were conducted with the Mould-$R$ and Gunn-$g$ filters (Figure \ref{scoperta}). Photometry (Table \ref{opt}) was performed relative to the SDSS $r$-band and $g$-band magnitudes of stars in the field \citep{York2000}, using our custom pipeline which performs image subtraction followed by PSF photometry on stacks of PTF images extracted from the PTF Infrared Processing and Analysis Center (IPAC) archive \citep[][]{Kate2012,Ofek2012,Ofek2013a}. The flux residuals from individual subtracted images were binned, and then converted to magnitudes. The errors were estimated from the standard deviation of the photometric measurements in each bin (Figure \ref{lc_mag}).

Multi-color optical ($gri$) optical light curves were also obtained using the Palomar 60-inch telescope \citep[P60;][]{Cenko2006} and the RATCAM optical imager on the robotic 2m Liverpool Telescope \citep[LT;][]{Steele2004} located at the Roque de Los Muchachos Observatory on La Palma. P60 and LT photometry was extracted using the same pipeline described above (Table \ref{opt}). 
\subsection{Spectral classification}
\label{Spectralclassification}
\begin{figure*}
\begin{center}
\includegraphics[width=12.cm]{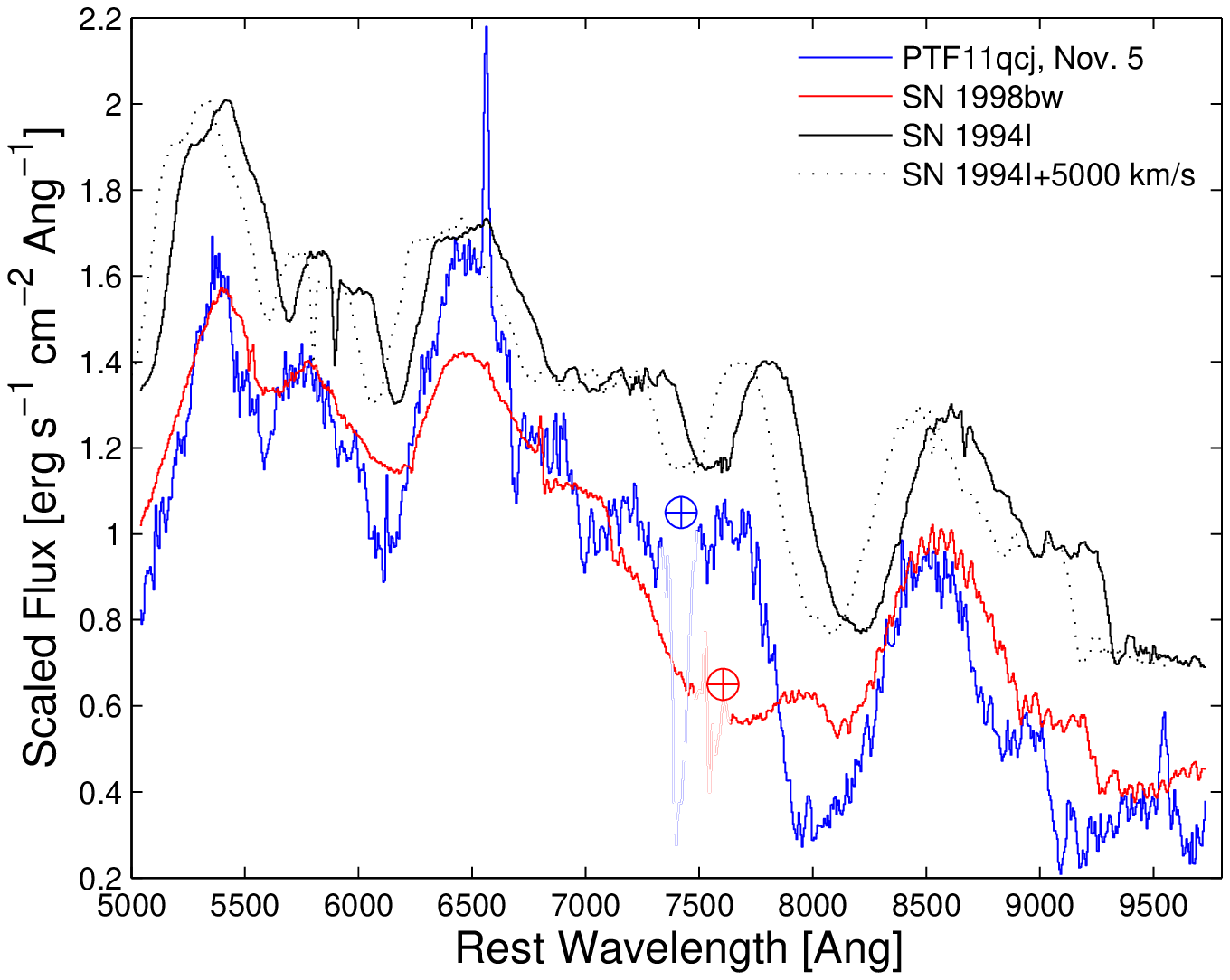}
\caption{Comparison of our first spectrum of PTF11qcj (blue line) with spectra of Type Ic SNe 1998bw \citep[Ic-BL at $\approx 19$ days post-maximum light, red line;][]{Patat2001} and 1994I \citep[normal type Ic at maximum light, black line; ][]{Clocchiatti1996}. Telluric absorptions have been excised. To match the spectrum of SN\,1994I, an additional blue shift of $\sim$5,000\,km\,s$^{-1}$ needs to be applied (dotted line). (See the electronic version of this paper for colors.) \label{spec}}
\end{center}
\end{figure*}

After the discovery of PTF11qcj with P48, we triggered a spectroscopic follow-up campaign\footnote{All spectra will be made public via WISeREP \citep{Yaron2012}} (Figure \ref{spec}). A first spectrum was taken on 2011 November 5 using OSIRIS on the Gran Telescopio Canarias\footnote{\url{http://www.gtc.iac.es/instruments/osiris/}} (GTC), with a 1\arcsec\  long-slit and the R300R grism ($\approx 5000-10000$\,\AA). The exposure time for this observation was $\approx 240$\,s. Two other spectra were obtained on 2011 November 7 and on 2011 December 21 using ISIS on the William Herschel Telescope\footnote{\url{http://www.ing.iac.es/Astronomy/instruments/isis/index.html}} (WHT), with a 1\arcsec\ long-slit, the R300B grating plus $\approx 4500$\,\AA\ central wavelength on the blue side; the R158R grating plus $\approx 7500$\,\AA\ central wavelength on the red side. The exposure times were of 600\,s and 1800\,s, respectively for the two observations. On 2011 November 26 and 2011 December 31, we obtained two spectra using LRIS \citep{LRIS} mounted on the Keck-I 10\,m telescope\footnote{\url{http://www2.keck.hawaii.edu/inst/lris/}}. For both spectra we used a 1\arcsec\ slit, with the 400/8500 grating plus $\approx 7850$\,\AA\ central wavelength on the red side, and the 400/3400 grism on the blue side. The exposure times were 1200\,s for the blue side and 720\,s for the red side, on 2011 November 26; 900\,s for the blue side and 840\,s for the red side, on 2011 December 31. Finally, on 2012 March 20, we took a last spectrum with DEIMOS \citep{DEIMOS} mounted on the Keck-II 10m telescope, using a 0.8\arcsec\ slit, and the 600ZD (600/7500) grism centered at $7000$\,\AA\,. The wavelength coverage was $4350-9650$\,\AA\,, the exposure time was $2\times 1200$\,s, and the spectral resolution was $\approx 3.4$\,\AA.

\begin{figure*}
\begin{center}
\includegraphics[width=12.cm]{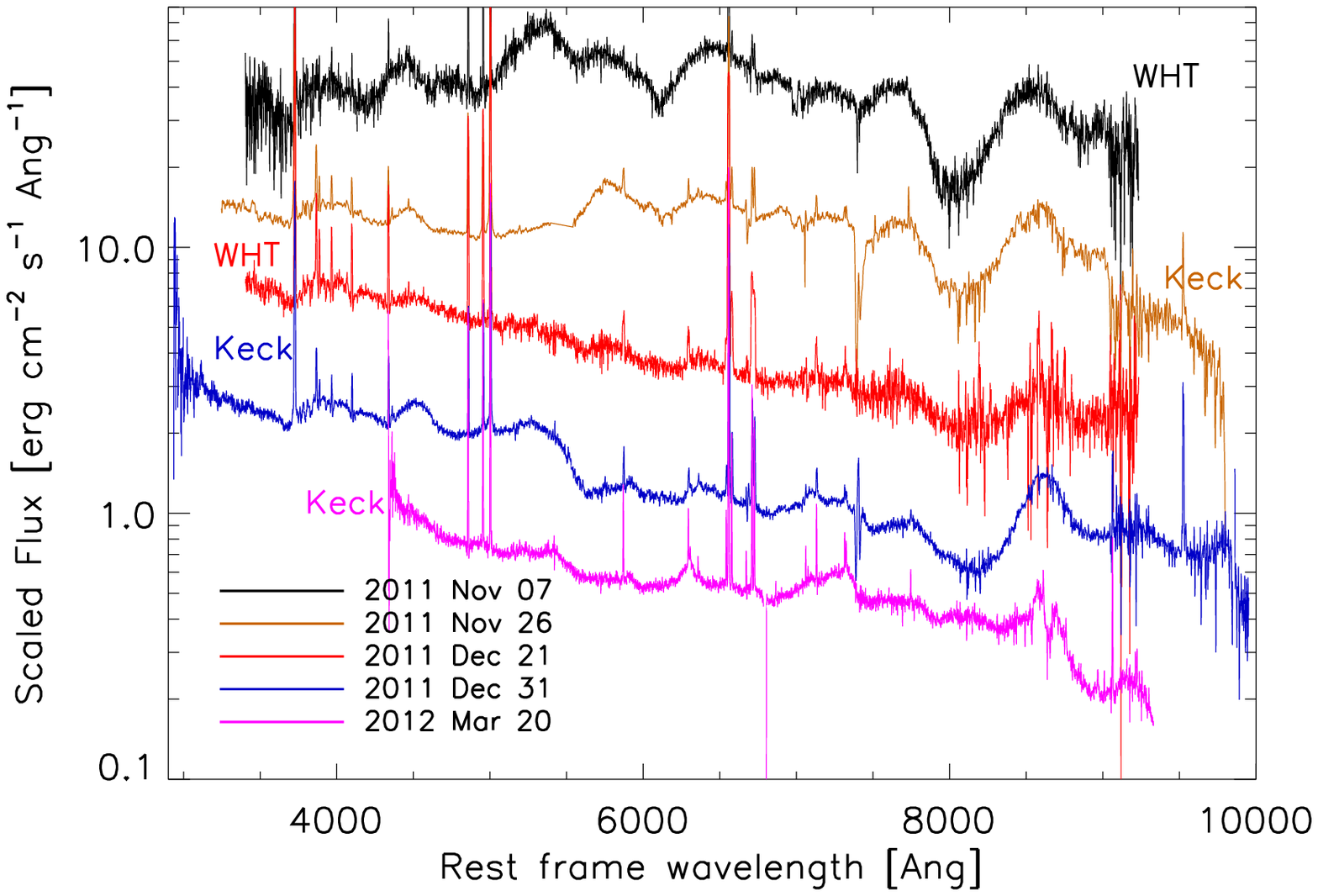}
\caption{Optical spectra of PTF11qcj obtained with the WHT (black and red) and Keck (orange, blue, and purple). From top to bottom, these spectra were taken on 2011 November 07 (black), 2011 November 26 (orange), 2011 December 21 (red), 2011 December 31 (blue), and 2012 March 20 (purple). See text for a detailed description of these spectra. (See the electronic version of this paper for colors.)\label{spec_all}}
\end{center}
\end{figure*}

Our first spectrum obtained on 2011 November 5 with the GTC (see Figure \ref{spec}) is found to be most similar to a type Ic SN by the Superfit software \citep{Howell2005}. A comparison with template spectra shows relatively high velocities (around 22,000\,km\,s$^{-1}$ derived from the Ca II IR triplet minimum), that are driving superfit to find the BL SN 1998bw (at $\approx 19$\,d post-maximum) as the best fit (Figure \ref{spec}, red line). However, more normal type Ic SN spectra (e.g., SN\,1994I, black line in Figure \ref{spec}) could also provide an acceptable fit, provided these are artificially blue shifted by $\sim$5,000\,km\,s$^{-1}$ (dotted line in Figure \ref{spec}). In fact, the absorption features profiles are more similar to SN\,1994I than to SN\,1998bw. Aside from nebular H$\alpha$ emission from the host galaxy, no strong narrow lines are seen in this first spectrum. 

The WHT spectrum taken two days later (Figure \ref{spec_all}, black line) extends to bluer wavelengths. No narrow lines are seen except for common host galaxy lines (Balmer series, O[III] and O[II]), and the continuum shape is red (with a prominent decreasing trend beyond $\approx 5500$\,\AA) as is common for SNe of type Ic. As the SN flux declines, most notably in the spectra taken on and after 2011 November 26, the SN signal is strongly mixed with underlying emission from the host galaxy. 

Given the classification as Ic-BL SN, we commenced, in addition to our optical spectroscopic follow-up, an extensive radio, X-ray, and IR follow-up campaign (Sections \ref{Radioobservations}-\ref{IRobservations}).
\subsection{Pre-explosion images and photometry}
We searched for evidence of pre-explosion activity in the PTF images taken before the discovery date of PTF11qcj. To this end, we constructed a reference frame from 30 images taken over the period 2013 March - May 2013 ($\approx 16$ months after the SN explosion, when PTF11qcj was no longer detected by P48). We geometrically resampled this reference image to the coordinate frame of every image of the PTF11qcj field taken before and during the SN explosion (hereafter, the ``science images''). We matched the PSFs of the resampled reference image and of the science images using a non-parametric kernel \citep[a process that usually results in the degradation of the reference image quality; e.g.,][]{Bramich2008}, and subtracted the reference image from each science image. Note that in this process the science frames are never geometrically resampled. PSF photometry was then performed on the difference images at the position of PTF\,11qcj, which was measured accurately from the difference images in which the SN is present, with the shape of the PSF measured from either the pre-subtraction science or reference images. This results in a flux and flux uncertainty computed at the position of PTF11qcj for every science image, regardless of whether the SN was present or not \citep[see also ][]{Ofek2013a}. 
\begin{figure*}
\begin{center}
\vbox{
\includegraphics[width=12cm]{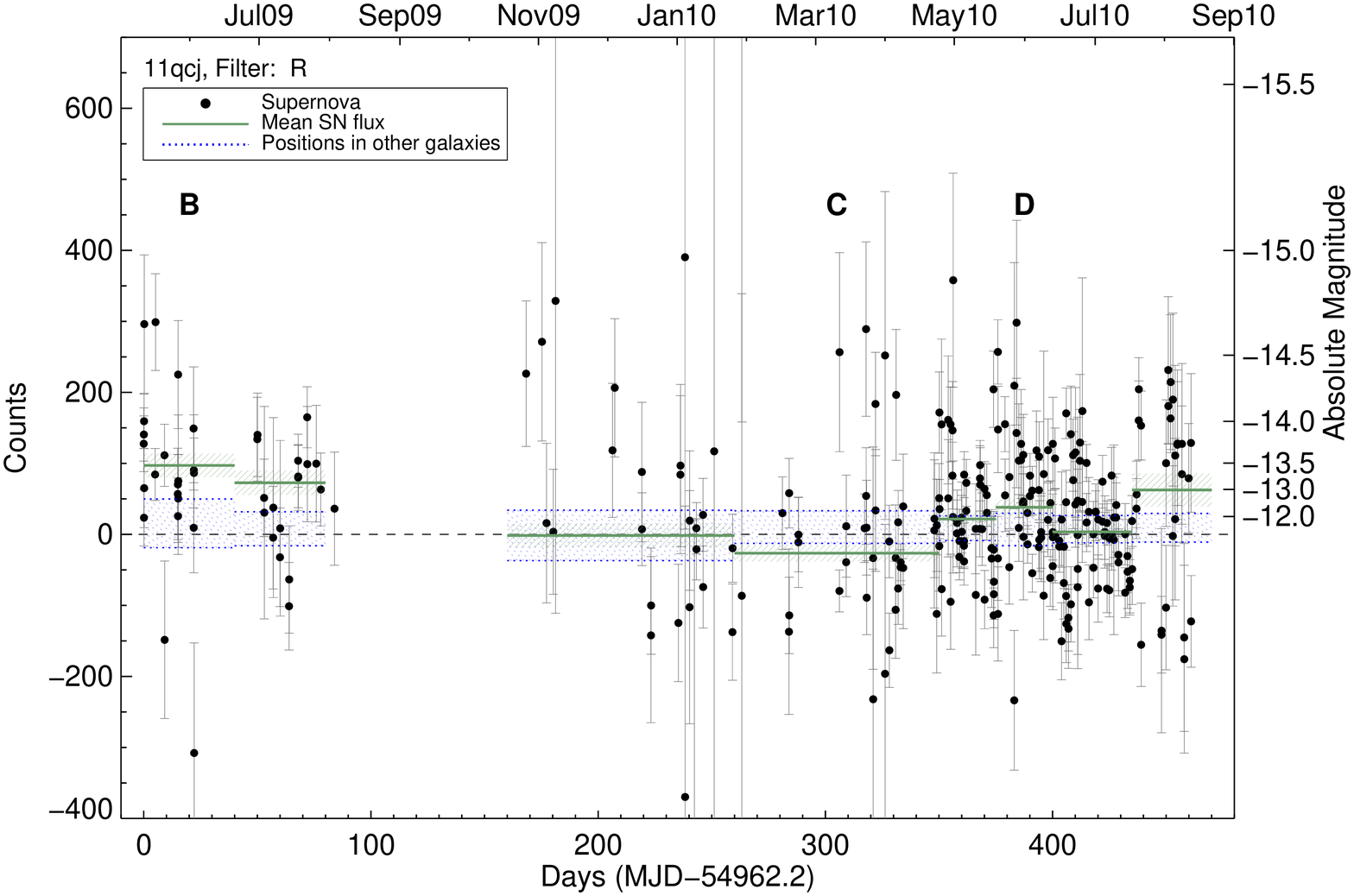}
\includegraphics[width=12cm]{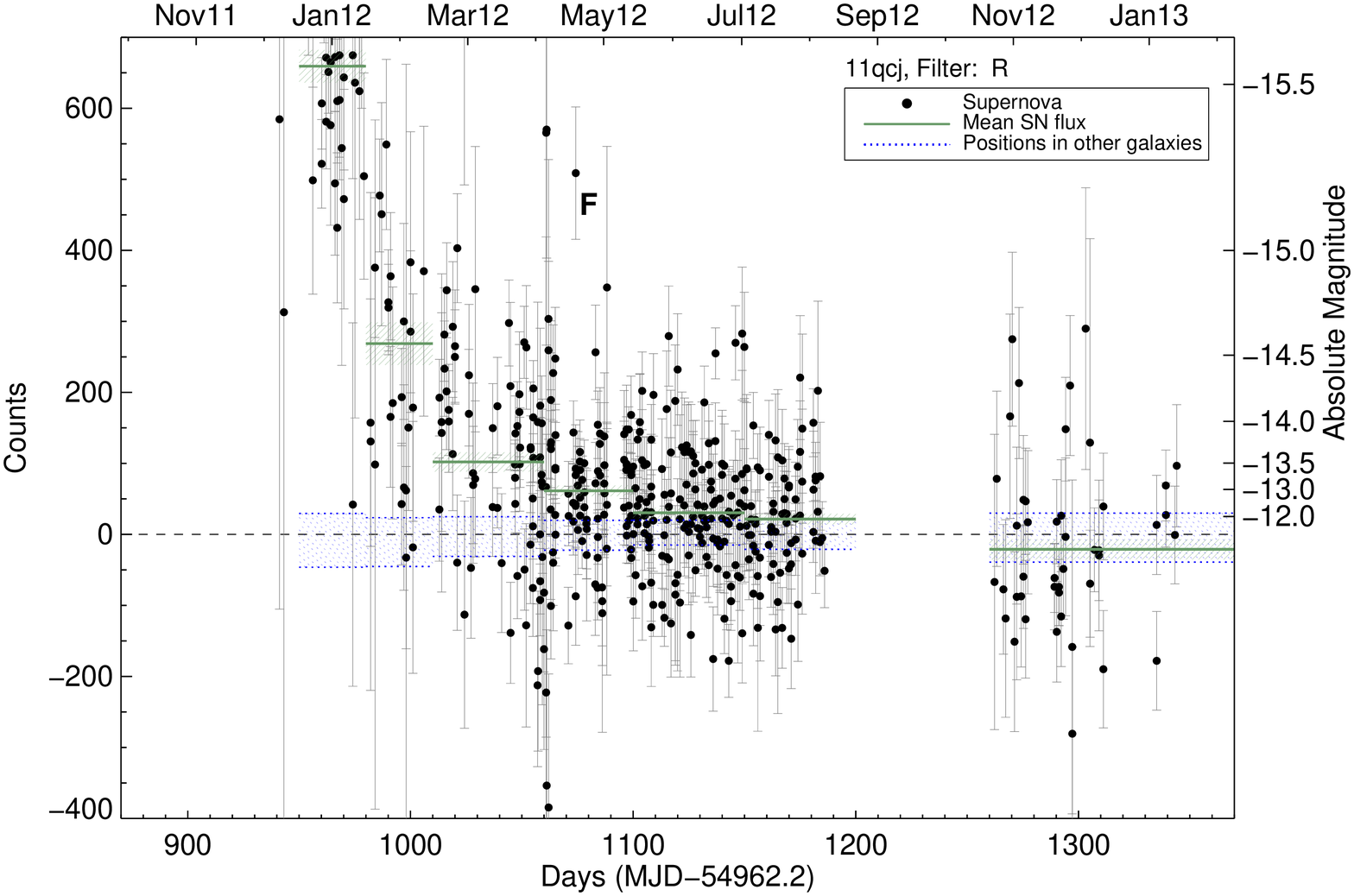}
}
\caption{The detailed P48 $R$-band light curve of PTF11qcj in flux space, with day zero defined as the first day on which PTF observed the field of PTF11qcj. Two seasons are shown. The first (top) shows the season 2\, yr before the SN explosion, and the second (bottom) shows the year after the SN explosion. The black circles show measurements made using PSF fitting at the SN position on the individual difference images. The green solid lines area shows the weighted-mean flux in time bins, and the extent of the green hashed area the error on the mean. The blue hashed area shows the typical range of PSF counts obtained by performing PSF photometry in other galaxies near to the host of PTF11qcj; this tracks any global image subtraction problems. The horizontal dashed line shows the position of zero flux. The letters B, C, D and F correspond to the photometry bins stacked to produce the images shown in Figure~\ref{preexplimg}. To convert the counts scale to apparent magnitude, use $m=-2.5\log(\mathrm{counts})+27$. (See the electronic version of this paper for colors.) \label{preexpllc}}
\end{center}
\end{figure*}

\begin{figure*}
\begin{center}
\includegraphics[width=12cm,angle=90]{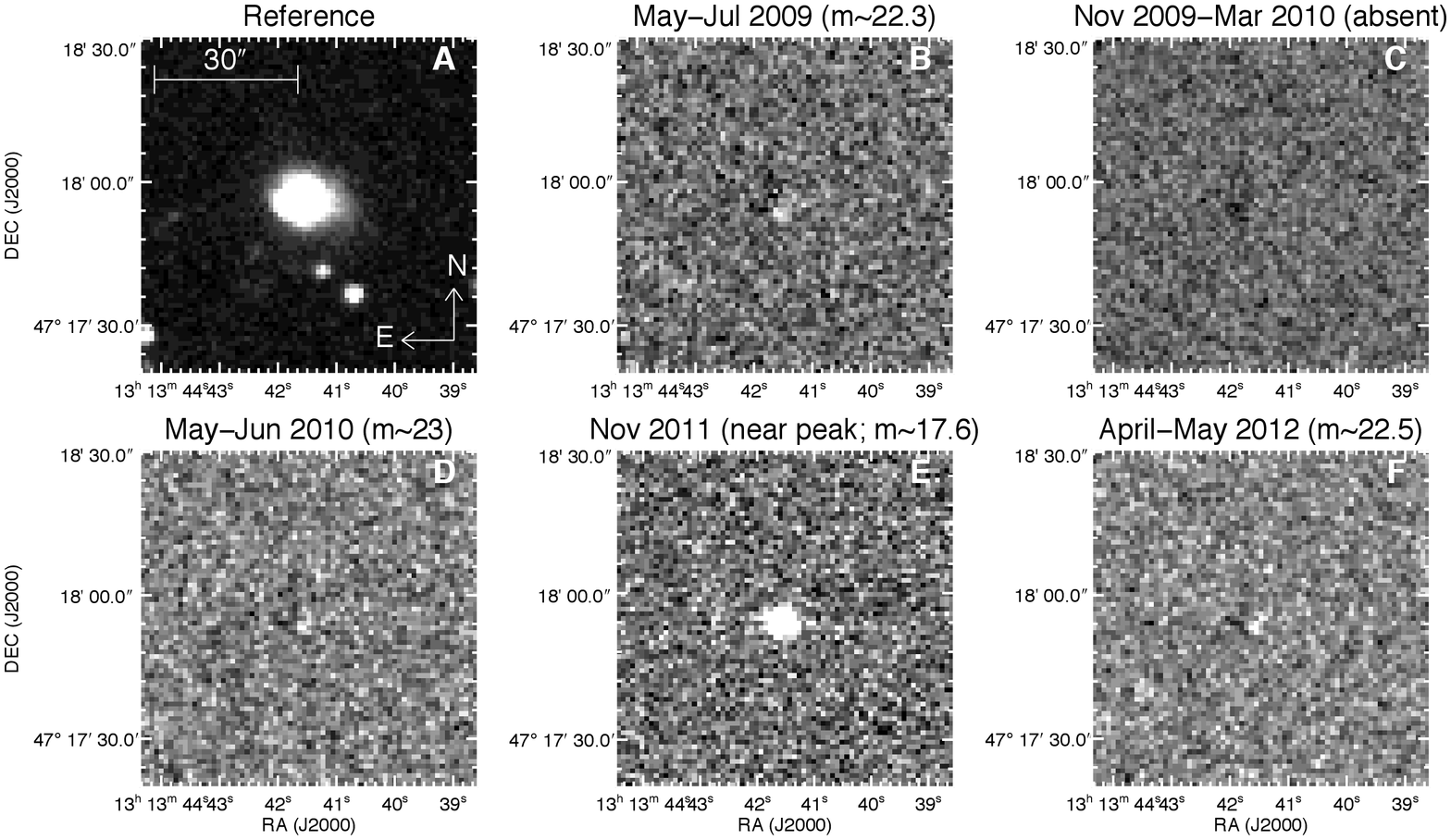}
\caption{Six postage stamps showing the area around PTF11qcj. Top left (panel A) shows a deep reference image generated from 30 images taken across March-May 2013. The remaining panels (B-F) show stacked image subtractions for the time periods indicated. Panels B-D are constructed from difference images from before the SN explosion, with an apparent faint precursor detection in panels B and D. Panel E is near the peak of the SN light curve, and panel F is $\sim$6 months after explosion once the SN had faded to a similar flux level as the candidate precursor detections. The positions of panels B, C, D and F in the SN light curve can be seen in Figure \ref{preexpllc}. See text for discussion. \label{preexplimg}}
\end{center}
\end{figure*}

In Figure \ref{preexpllc} we show the light curve data obtained for the pre-SN season of May 2009 - September 2010 (top panel), and for the post-SN season of January 2012 - January 2013 (bottom panel). In each of these Figures, the black circles show measurements made using PSF fitting at the SN position on the individual difference images; the green area shows the weighted-mean flux in time bins, and the extent of the error on the mean; finally, the blue area shows the typical range of PSF counts obtained by performing PSF photometry on other galaxies in the PTF11qcj field, having similar magnitude as the PTF11qcj host ($\pm$2\,mag), and at the same isophotal radius as the one at which PTF11qcj \textbf{occurred in its own host}. Thus, while the blue areas in these Figures track any global image subtraction problems, the green areas track departures in the mean flux from zero and allow us to asses the presence of any pre-SN activity. 

As evident from Figure \ref{preexpllc}, during the pre-SN season (top panel) we find tentative evidence for periods of activities in between May 2009 and July 2009 (see also panel B in Figure \ref{preexplimg}), and between May 2010 and June 2010 (see also panel D in Figure \ref{preexplimg}). These active periods are separated by a period of non-detection, between November 2009 and May 2010 (see also panel C in Figure \ref{preexplimg}). Specifically, we note that during the active period of May 2009 - July 2009, the possible precursor from PTF11qcj is as bright as the SN was during April 2012 - May 2012 (this is also evident comparing panels B and F in Figure \ref{preexplimg}). We finally checked whether any evidence for excess flux was still present during the post-SN season (after the SN itself faded below the P48 detection limit). As evident from the bottom panel of  Figure \ref{preexpllc}, at $t\gtrsim 1$\,yr after explosion no excess flux is detected at the position of PTF11qcj. This further supports the evidence for precursor activity during the pre-SN season. 
\subsection{Radio observations}
\label{Radioobservations}
On 2011 November 15, we started a long-term monitoring campaign of PTF11qcj (along with calibrators J1327+4326 and 3C\,286) with the Karl G. Jansky Very Large Array\footnote{The National Radio Astronomy Observatory is a facility of the National Science Foundation operated under cooperative agreement by Associated Universities, Inc.; \url{https://public.nrao.edu/telescopes/vla}} \citep[VLA;][]{Perley2009} in its D, DnC, C, CnB, and A configurations, under our Target of Opportunity programs\footnote{VLA/11A-227, VLA/11B-034, VLA/11B-247, VLA/12B-195 - PI: A. Corsi}. VLA data were reduced and imaged using the Common Astronomy Software Applications (CASA) package. 

The light curves of PTF11qcj at frequencies of $2.5$\,GHz, $3.5$\,GHz, $5$\,GHz, $7.4$\,GHz, $13.5$\,GHz, $16$\,GHz are reported in Table \ref{radioTab}. The VLA measurement errors are a combination of the rms map error which measures the contribution of small unresolved fluctuations in the background emission and random map fluctuations due to receiver noise, and a basic fractional error (here estimated to be of $\approx 5\%$) which accounts for inaccuracies of the flux density calibration \citep[see e.g.,][]{Weiler1986,Eran2011}. 

We also observed the field of PTF11qcj (together with the test calibrator J1203+480) using the Combined Array for Research in Millimeter-wave Astronomy\footnote{\url{http://www.mmarray.org/}}  (CARMA), at a frequency of $93$\,GHz. Bad weather caused de-coherence which resulted in a limited follow-up of this source with CARMA. Nevertheless, the data collected on 2011 November 19 and 2011 November 26\footnote{CARMA program \#c0857; PI: A. Horesh} both resulted in a detection of PTF11qcj (Table \ref{radioTab}).
\subsection{X-ray observations}
\label{X-rayobservations}
We have carried out an X-ray monitoring campaign of PTF11qcj with \textit{Chandra}\footnote{\url{http://www.nasa.gov/mission_pages/chandra/main/index.html}} \citep{Weiss2002} and \textit{Swift}\footnote{\url{http://heasarc.nasa.gov/docs/swift/}} \citep{Gehrels2004}. All our \textit{Swift}-XRT observations yielded non detections (see Table \ref{X}), while \textit{Chandra} detected PTF11qcj in three epochs\footnote{DDT proposals \#501793, \#501794, \#501797; PI: A. Corsi}. The results of our X-ray follow-up are reported in Table \ref{X}. 

We reduced \textit{Swift} data adopting an on-source circular aperture of $7.2$\,\arcsec and assuming that 37\% of the photons are within this aperture. In addition to the results obtained for single-epoch observations (see Table \ref{X}), we also provide the upper-limit obtained by co-adding all of the \textit{Swift} non-detections. 

For \textit{Chandra} observations, we adopted a circular aperture of $2''$ fixed on the optical position of PTF11qcj. The background has been estimated from nearby portions of the image. Net source counts have been derived by subtracting from the counts measured in the on-source region the number of counts measured in the background region, scaled by the ratio of the on-source to background region areas. We caution that the background estimated from nearby portions of the image does not include any diffuse or point sources associated with the host galaxy itself. In this sense, the count rates indicated in Table \ref{X} are upper-limits on the flux from PTF11qcj. Since the first epoch is brighter than the third, we can however be confident that we have seen some emission from PTF11qcj during the first epoch. 
\subsection{IR observations}
\label{IRobservations}
\begin{figure*}[!h]
\begin{center}
\includegraphics[width=8cm,angle=-90]{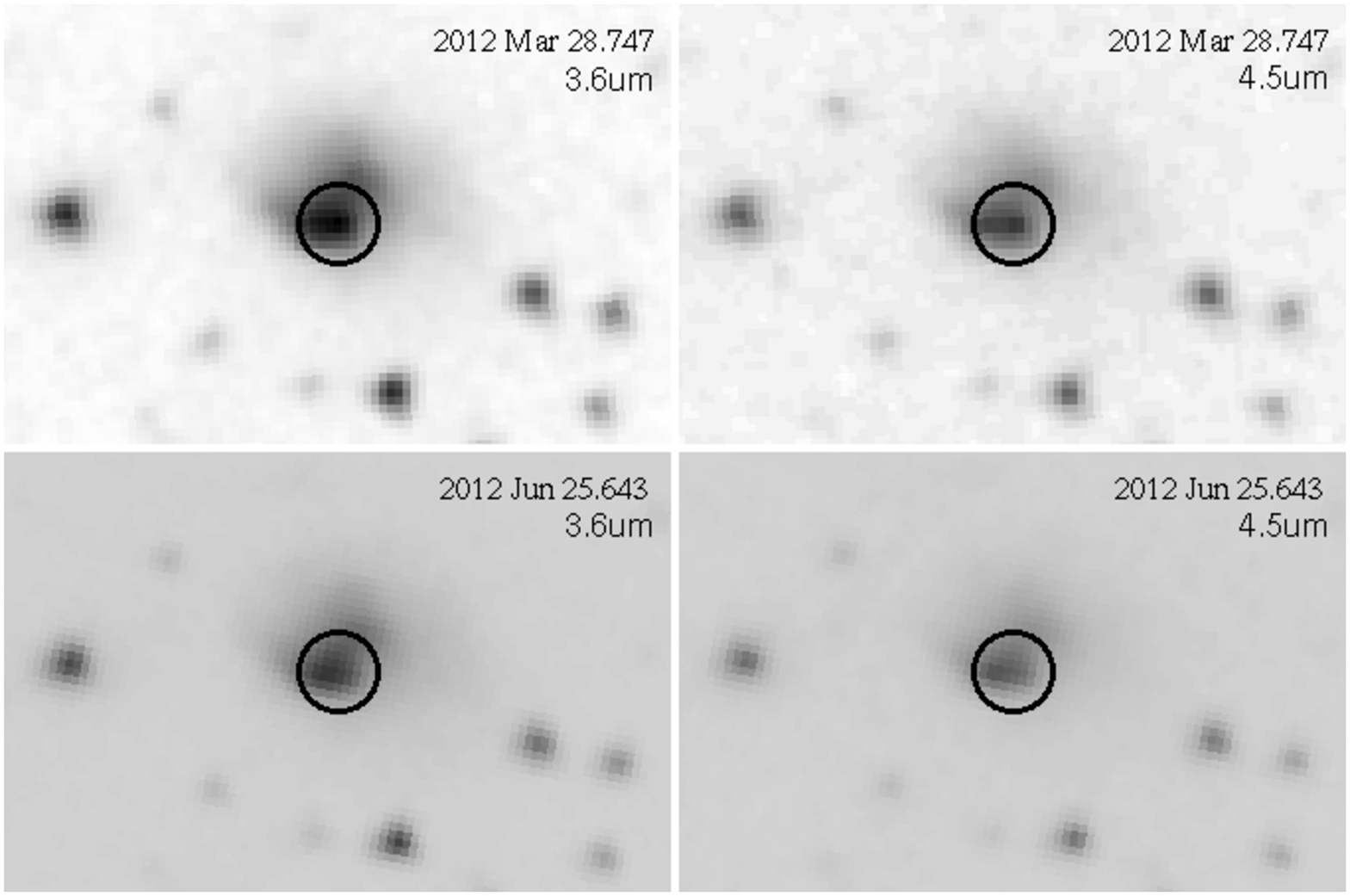}
\caption{\textit{Spitzer} images of the PTF11qcj field. The black circle indicates the position of PTF11qcj. Contamination from the host galaxy is evident in all the images.\label{spitzer}}
\end{center}
\end{figure*}

We observed the position of PTF11qcj with \textit{Spitzer}\footnote{\url{http://www.spitzer.caltech.edu/}} \citep{Fazio2004} on two epochs\footnote{DDT proposal \#31731; PI: A. Corsi} (on 2012 March 28.747 and 2012 June 25.643; Table \ref{spitzertab}). Data were reduced using the standard \textit{Spitzer}/IRAC pipeline. Subsequently, aperture photometry was performed using a radius of 4 pixels (i.e., $2.4\arcsec$) for PTF11qcj, and an annulus of radii 8 to 16 pixels (i.e., $4.8\arcsec$ to $9.6\arcsec$)for the sky background. Calibration and aperture corrections were as per the zero points listed in the \textit{Spitzer}/IRAC handbook. However, there is clearly a host galaxy contribution at the position of PTF11qcj (see Figure \ref{spitzer}): for this reason, we consider the flux measurement obtained with \textit{Spitzer} as an upper-limit to the IR flux of PTF11qcj.

On 2012 March 28, we also observed the field of PTF11qcj in $K_s$-band with the Palomar 200-inch telescope (P200). The contribution of the host galaxy is clearly evident at the position of PTF11qcj in the P200 image (Figure \ref{scoperta}, lower-right panel). Performing aperture photometry with respect to two reference stars in the field (2MASS 13134736+4719100 and 13135239+4717152), we derive a $K_s$-band photometric data point that, as for the \textit{Spitzer} observations, we consider as an upper-limit to the SN flux in this band (Table \ref{spitzertab}).
\section{Multi-wavelength analysis}
\label{Analysis}
\subsection{Spectroscopic properties}
\label{spectralproperties}
As discussed in Section \ref{Spectralclassification}, our first spectrum of PTF11qcj shows relatively high velocities (around 22,000\,km\,s$^{-1}$). The spectral similarity between PTF11qcj at $\approx 13$\,d after $g$-band discovery (2011 November 5, or 55870\,MJD), and the Ic-BL SN\,1998bw at $\approx 19$\,d after peak (Figure \ref{spec}), suggests that we discovered PTF11qcj around peak time or shortly thereafter. We thus infer an explosion date of 55830\,MJD\,$\lesssim t_e\lesssim 55850$\,MJD, assuming an explosion-to-peak timescale of $\approx 7-20$\,d, as derived from well studied Ib/c SN samples \citep[e.g., Figure 5 in][]{Drout2011}. 
\begin{figure*}
\begin{center}
\includegraphics[width=14.cm]{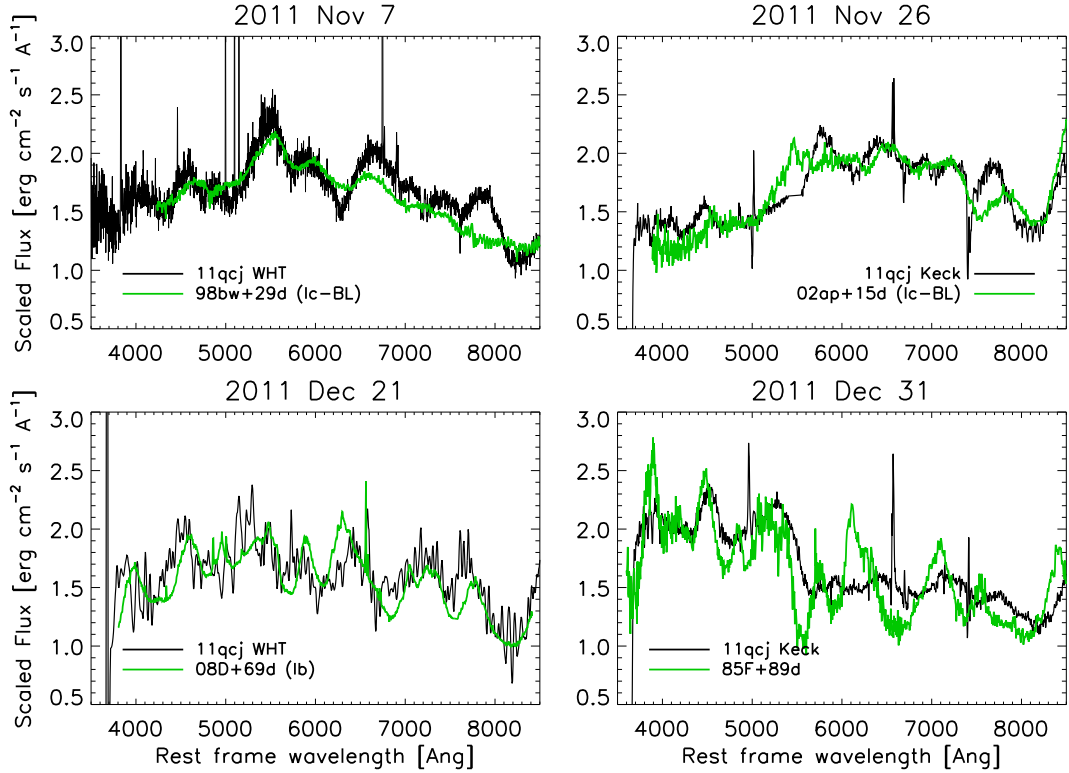}
\caption{Optical spectra of PTF11qcj (black) compared with the spectra of other notable SNe (green). All the spectra except for the upper-left one have been host subtracted as described in the text. The earlier spectra (upper panels) suggest a match with SNe of the Ic-BL type \citep[SN\,1998bw and SN\,2002ap; ][]{Galama1998,Mazzali2002,galyam2002}. The later spectra (lower panels) are better matched by more normal SNe Ib/c \citep[SN\,2008D and SN\,1985F; ][]{Filippenko1986,Begelman1986,Mazzali2008,Soderberg2008}. (See the electronic version of this paper for colors.) \label{spec_new}}
\end{center}
\end{figure*}

\begin{figure}
\begin{center}
\includegraphics[width=9.cm]{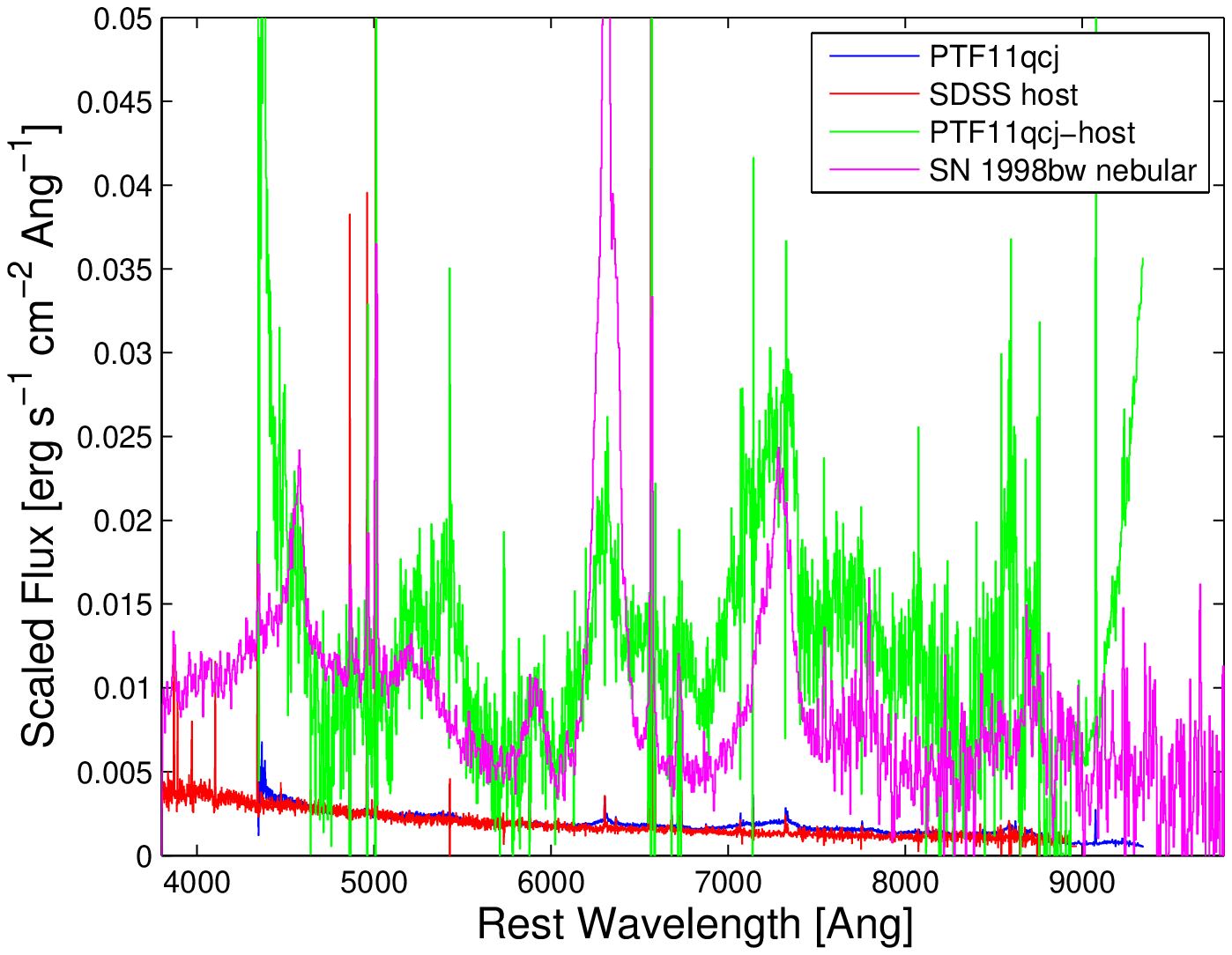}
\caption{Comparison between our last spectrum of PTF11qcj (taken on 2012 March 20) and the nebular spectrum of SN\,1998bw. (See the electronic version of this paper for colors.) \label{spec_nebular}}
\end{center}
\end{figure}

The continuum shape observed in the spectrum of PTF11qcj taken on 2011 November 26 (Figure \ref{spec_all}, orange line), is flat or even rising toward the blue side down to $\approx 3500$\,\AA. Indeed, this spectrum and the ones taken after 2011 November 26 show the SN signal strongly mixed with underlying emission from the host galaxy. We have attempted to subtract this contribution using a spectrum of the host galaxy obtained by the SDSS survey\footnote{\tiny \url{ http://skyserver.sdss3.org/dr9/en/get/specById.asp?id=1642855417023326208}}. We scaled the host spectrum until the strong emission lines approximately matched in intensity the lines observed in our SN spectra, rebinning the spectra as needed to match the line width, and then subtracted the host spectrum (resampled onto the SN spectrum wavelength scale) from our observed spectra. These host-subtracted spectra are plotted in Figure \ref{spec_new}. 

We run the Superfit program on the host-subtracted spectra and find a reasonable match to spectra of SNe Ib/c of comparable ages. The earlier-epoch spectra seem to fit SNe of the Ic-BL class, while later-epoch ones are better matched by more normal SNe Ib/c. We stress that the host-subtraction procedure we attempted is quite rough due to the irregular nature of the host and an apparent bright star-forming knot seen in pre-explosion SDSS imaging, co-located with the SN. Still, in view of the matches found between the host-subtracted spectra and observations of other events, it is certainly reasonable to assume that PTF11qcj evolved in a similar way to other SNe Ib/c.

Finally, the late-time spectrum (2012 March 20; Figure \ref{spec_all}, purple line) is nebular, with O\,I 6300\,\AA\ and Ca\,II emission lines. After subtracting the host contribution, this spectrum appears quite similar to late spectra of SN\,1998bw, again within the large uncertainties related to host contamination (Figure \ref{spec_nebular}).

In order to estimate the host-galaxy extinction toward the SN, we measured the Na D absorption doublet in our high signal-to-noise and high-resolution Keck/DEIMOS spectrum obtained on March 20, 2012. We detect weak features consistent with the Na D doublet at the host redshift, for which we measure the following equivalent width values using the IRAF/splot routine: EW$_{5890 \rm \AA}=0.087$\AA, and EW$_{5896 \rm \AA}=0.265$\AA. Since the 5890\AA\, component is always stronger (opposite what we see), and the FWHM of the 5896\AA\, component is almost 3 times larger than that of the 5890\AA\, line, we assume that the 5896\AA\, doublet component is probably contaminated by a noise fluctuation. Using the relation of \citet{Poznanski2012} (see their Figure 7) we estimate an extinction of  E$_{\rm B-V}$=0.03 (A$_{\rm R}\sim0.08$\,mag) using the 5890\AA\, component.  We thus apply this small correction, together with the correction for Galactic extinction, to our photometry plotted in Figure \ref{lc_mag}.
\subsection{Pre-SN activity: a precursor from PTF11qcj progenitor?}
\label{LBVburst}
In recent years pre-SN eruptions have been detected from type IIn SNe \citep[e.g.,][]{Mauerhan2013,Prieto2013,Ofek2013a} and type Ibn SNe \citep[e.g.,][]{Nakano2006,Pastorello2007,Foley2007}. The total ejected mass in these outbursts is estimated to be a fraction of a solar mass \citep[e.g.,][]{Immler2008,Ofek2013b,Ofek2013a}. However, pre-SN eruptions were never detected before from other types of SNe.

Our pre-explosion images of PTF11qcj show tentative evidence for the existence of pre-explosion activity, particularly in terms of a precursor event reaching $m_r\approx 22.3$\,mag during May-July 2009, $\approx 2.5$\,yrs before the SN discovery. Marginal evidence for further pre-SN activity ($m_r\approx 23$\,mag) is found during May-June 2010, $\approx 1.5$\,yrs before discovery. 

At the distance of PTF11qcj ($d_L\cong 124$\,Mpc), correcting for local (see Section \ref{spectralproperties}) and Galactic \citep[$A_r=0.024$\,mag;][]{Schlafly2011} extinction at the SN position, $m_r\approx 22.3$\,mag ($m_r\approx 23$\,mag) corresponds to $M_r\approx -13.3$\,mag ($M_r\approx -12.6$\,mag). For comparison, the precursor observed 2\,yr before explosion of the prototype Ibn SN\,2006jc reached $M_r\approx -14$\,mag \citep{Pastorello2007}. Based on our pre-SN images, we can exclude a precursor as bright as the one observed for SN\,2006jc during the epochs covered by our observations (see Figure \ref{preexpllc}).
\subsection{SN optical light curve analysis}
\label{Opticallightcurve}
As discussed in Section \ref{spectralproperties}, we likely discovered PTF11qcj around peak time. Based on our first detection with P48 ($m_r\approx 17.6$\,mag), we can constrain the peak $r$-band magnitude to be $m_{\rm r, peak}\lesssim 17.6$\,mag. Correcting for extinction (Galactic plus local), our upper-limit on the peak $r$-band magnitude of PTF11qcj ($m_{\rm r, peak}\lesssim 17.6$\,mag) corresponds to an absolute $r$-band peak magnitude of $M_{\rm r,peak}\lesssim-18$\,mag. This limit is compatible with the average $R$-band peak magnitudes found in recent systematic studies of Ib/c SNe \citep{Drout2011}. 

In the 15 days following the P48 discovery, the P48 $r$-band light curve of PTF11qcj decreased by $\Delta m_{r}\approx 0.9$\,mag. The optical light curve of the GRB-associated SN\,1998bw dropped by 1.1\,mag in the first 15\,d after peak \citep{McKenzie1999}. Note that for PTF11qcj the 15-days magnitude drop is calculated with respect to the discovery time rather than the peak time; however, we likely discovered the PTF11qcj around peak. In Figure \ref{lc_mag}, we show a comparison between the PTF11qcj $r$-band light curve, and the light curves of the Ic SN\,1994I \citep{Richmond1996}, and the Ic-BL SN\,1998bw \citep{Clocchiatti2011}, scaled to the distance of PTF11qcj. The explosion time of PTF11qcj is taken as $\approx 15$\,d before peak, where the last is assumed to be coincident with our $g$-band discovery (as suggested by our spectral analysis). A rise time of $\approx 15$\,d is a factor of $\approx 1.3$ shorter than the rise time of SN\,1998bw ($\approx 20$\,d). Indeed, the SN\,1998bw $R$-band light curve compressed by a factor of $\approx 1.33$ (Figure \ref{lc_mag}, orange dashed-line) seems to match the one of PTF11qcj.  
 
The early-time ($t\lesssim 60$\,d) evolution of Ib/c SNe is usually referred to as the photospheric phase. This phase is characterized by a high optical depth, and a post-maximum light curve decay rate slower than the $^{56}$Ni-to-$^{56}$Co decay. During this phase, we can use the light curve properties to estimate the physical parameters of the SN explosion \citep{Arnett1982}: the $^{56}$Ni mass, the ejecta mass, and the kinetic energy. From Figure 22 in \citet{Drout2011}, for $\Delta m_{15,r}\approx 0.9$\,mag and $M_{\rm r, peak}\lesssim -18$\,mag, we roughly estimate $M_{\rm Ni}\gtrsim 0.1\,M_{\odot}$ and $(E_{K}/10^{51}\,{\rm erg})^{-1/4}(M_{\rm eje}/M_{\odot})^{3/4} \approx 1.2$. For a photospheric velocity of 22,000\,km\,s$^{-1}$ (Section \ref{spectralproperties}), these values imply \citep[e.g., Equations (1)-(2) in][]{Drout2011} $M_{\rm eje}\approx 2.5\,M_{\odot}$ and $E_{K}\approx 8\times 10^{51}$\,erg. While these are very tentative estimates due to the uncertainties in the PTF11qcj peak-time and magnitude, these values are compatible with the ones observed for other Ib/c SNe. Particularly, the energy is consistent with what observed for BL-Ic and engine-driven events \citep{Drout2011}.
\subsection{Radio emission modeling}
\label{radiomodel}
\begin{figure*}
\begin{center}
\includegraphics[width=14.cm]{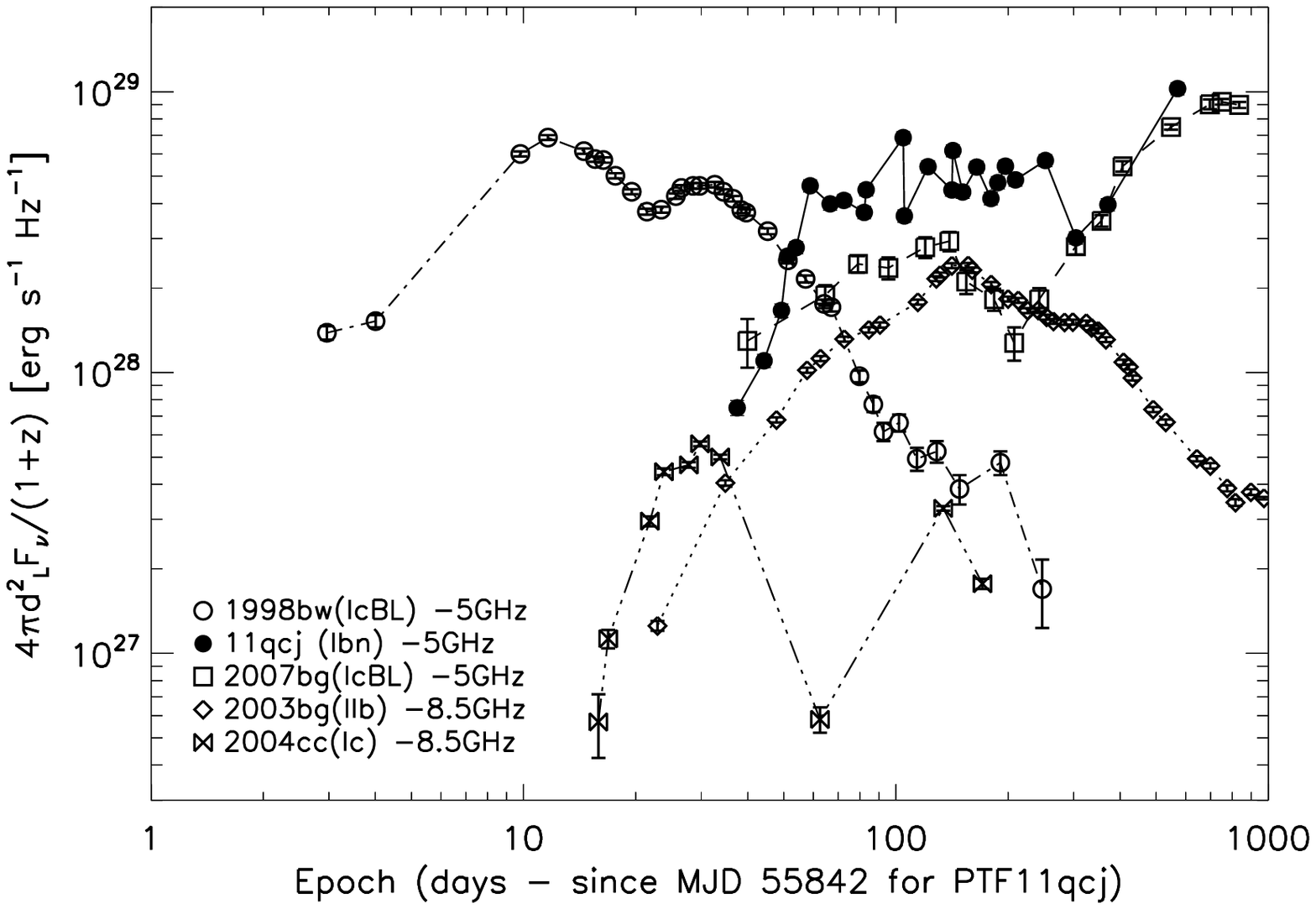}
\caption{Radio light curves of Ibc/IIb supernovae (5\,GHz or 8.5\,GHz) compared with the 5\,GHz light curve of of PTF11qcj. The GRB-associated SN\,1998bw (open circles) is peculiar for its bright radio emission and early-time peak (which is indicative of relativistic expansion). PTF11qcj (filled circles) radio luminosity is comparable to SN\,1998bw, but the peak time is shifted to longer timescales. SN\,2003bg (diamonds) was a IIb SN showing abrupt achromatic flux variations likely associated to episodic density enhancements \citep{Soderberg2003bg}. Long-term flux variations are also evident in the case of the Ic SN\,2004cc \citep[bow ties;][]{Wellons2012}. PTF11qcj shows even more drastic light curve variations, with the most abrupt being a variation of a factor of $\approx 2$ in flux over a timescale of $1$\,d around epoch $\approx 100$\,d since 55842\,MJD (we caution, however, that this abrupt variation is seen in data collected while the VLA antennas were being moved). SN\,2007bg \citep[squares;][]{Salas2012} is an example of Ic-BL SN with complex circumstellar environment and a late-time re-brightening similar to the one observed for PTF11qcj. \label{radio_comparison}}
\end{center}
\end{figure*}

Our radio monitoring campaign reveals that PTF11qcj is among the most radio-luminous Ib/c SNe. With a peak luminosity of $L_{\rm 5 GHz}\approx 10^{29}$ erg\,s$^{-1}$\,Hz$^{-1}$, PTF11qcj is comparable to the GRB-associated SN\,1998bw \citep[Figure \ref{radio_comparison};][]{Kulkarni1998}. From the presence of incoherent radio emission, assuming a non-relativistic source, we can derive a lower-limit on the size of the emitter ($\Theta$), by imposing that the brightness temperature $T_B$ does not exceed the equipartition value of $T_{B,eq}\approx 10^{11}$\,K \citep{Readhead1994,Kulkarni1998}:
\begin{equation}
T_{B}=\frac{c^2f_\nu}{2k_B\nu^2\pi\Theta^2}\lesssim 10^{11}\,{\rm K}\Rightarrow\Theta\gtrsim 2.1 \left(\frac{f_{\nu}}{\mu{\rm Jy}}\right)^{1/2}\left(\frac{\nu}{\rm GHz}\right)^{-1}\,\mu{\rm as}
\label{equi}
\end{equation} 
where $k_B$ is Boltzmann's constant, $\nu$ is the observing frequency, $f_{\nu}$ is the observed flux density, and $\Theta$ is the angular diameter of the emitting region. As evident from the above Equation, the largest lower limits are obtained for the lowest frequencies (and highest fluxes). Adopting the flux measured at $2.5$\,GHz around day 100 (when the low-frequency radio flux is around peak), we get $\Theta_{\rm 100\,d}\gtrsim 33\,\mu$as. At the distance of PTF11qcj, this corresponds to a size of $2 r \gtrsim 6\times 10^{16}$\,cm, or an average expansion speed $r/{\rm 100\,d}\gtrsim 0.1$\,c. We note that typical average speeds measured for the fastest moving ejecta of non GRB-associated Ib/c SNe are in the range $(0.03-0.3)c$ \citep{Berger2003,Wellons2012}.

A remarkable feature of PTF11qcj is that its radio light curves do not show a smooth evolution (see Figure \ref{radio_comparison}). This behavior, while difficult to model, is not completely unusual. Indeed, \citet{Soderberg2006} found that  $\approx 50\%$ of radio SNe show evidence for abrupt light curve variability, including abrupt steepening, abrupt rise, or episodic variations ($\lesssim$ than a factor of 2 in flux) of the radio light curve. These are typically explained as inhomogeneities in the CSM. We caution, however, that data corresponding to the most abrupt flux variation observed for PTF\,11qcj - a factor of $\approx 2$ in about 1\,day between $\approx 55949.6$ and $\approx 55950.4$\,MJD - were collected in 2012 while the VLA antennas were being moved from the D to the C configuration.
 
\begin{figure*}
\begin{center}
\includegraphics[width=14.cm]{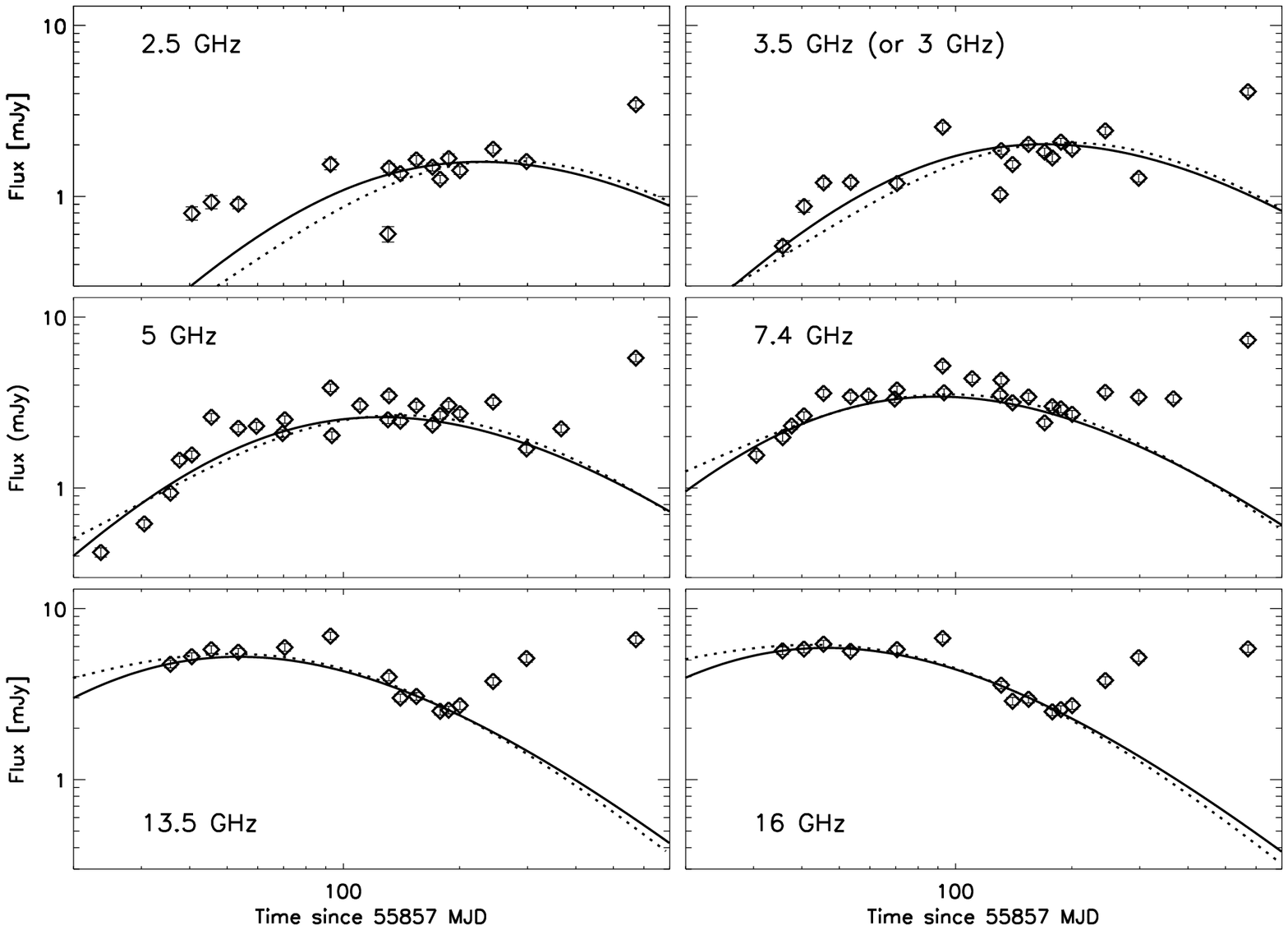}
\vspace{0.5cm}
\caption{Radio light curves of PTF11qcj obtained with the VLA. Synchrotron self-absorbed models with parameters as described in the text are also shown for comparison (solid and dotted lines). We note that the models break down during our latest three multi-frequency observations of this source, when a radio re-brightening is evident (especially at the optically thin frequencies). \label{radiolc}}
\end{center}
\end{figure*}

Radio SNe are known to emit non-thermal synchrotron radiation. According to the model proposed by \citet{Chevalier1982}, the relativistic electrons and amplified magnetic fields necessary for synchrotron emission are produced in the SN blast wave during the interaction with an ionized CSM. The last is usually assumed to be emitted via a constant mass-loss rate, constant velocity wind (i.e., $\rho_{\rm CSM}= \dot{M_w}/(4\pi v_wr^2)$) from a massive progenitor. The observed radio emission is characterized by a smooth turn-on first at higher frequencies, and later at lower frequencies, usually explained as temporally decreasing self-absorption as the shock propagates toward lower density regions. The initial turn on may also be related to free-free absorption in the ionized CSM \citep[][]{Chevalier1998}. While external free-free absorption from a homogeneous medium is predicted to cause an exponential rise of the low-frequency radio flux, a power-law is characteristic of self-absorption, internal absorption from thermal absorbing gas mixed into the synchrotron emitting gas, or free-free absorption from a clumpy external medium \citep{Weiler1990}. 

\begin{figure*}
\begin{center}
\includegraphics[angle=90,width=12.cm]{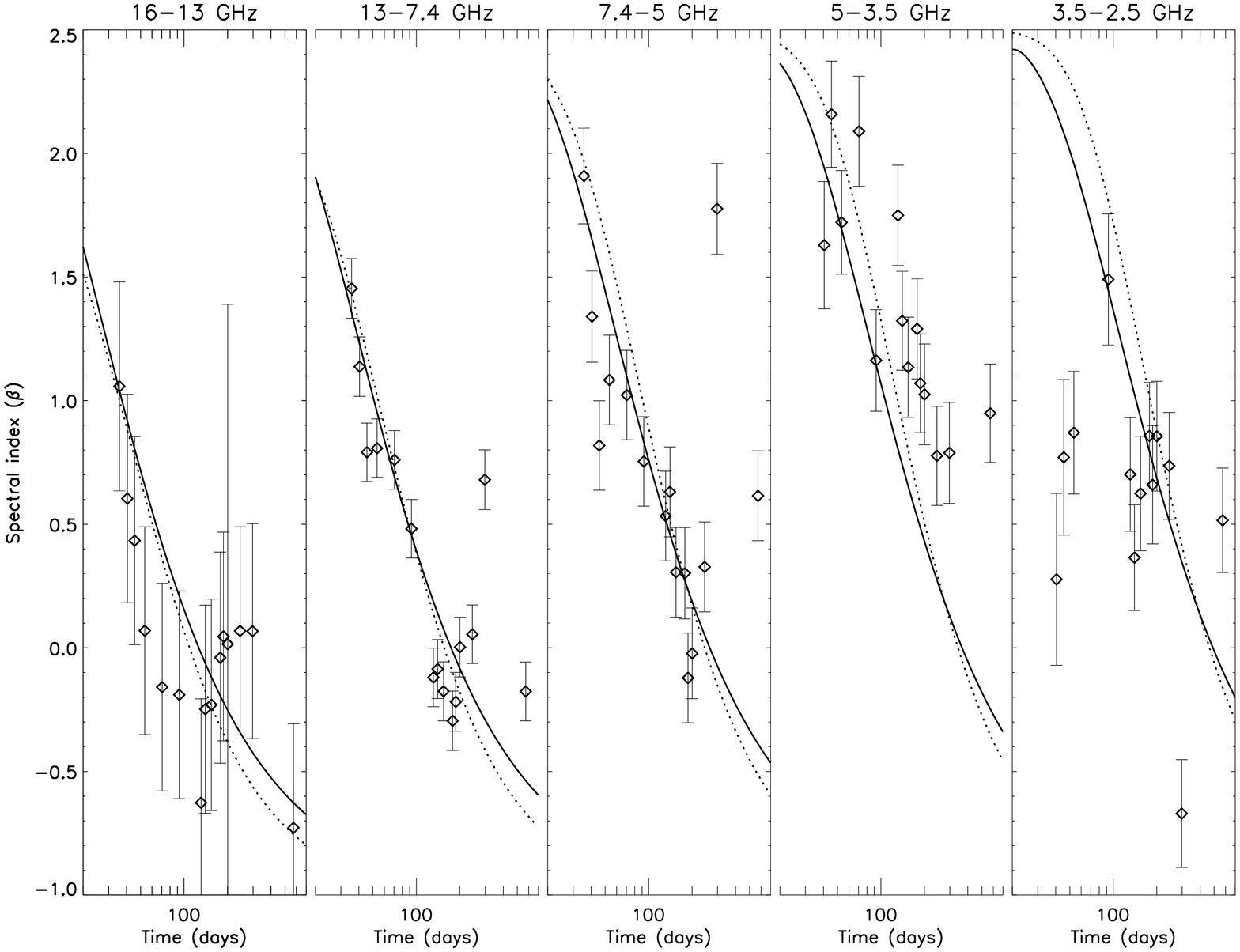}
\caption{Observed evolution of PTF11qcj radio spectral indices $\beta$ (with $f_{\nu}\propto \nu^{\beta}$). Time is calculated in days since time $t_e=55857$\,MJD. The spectral index evolution as predicted by the radio models discussed in the text is also shown for comparison (solid and dotted lines). \label{spectralindex}}
\end{center}
\end{figure*}

In Figure \ref{spectralindex} we show the observed spectral indices $\beta$ (with $f_{\nu}\propto \nu^{\beta}$) for PTF11qcj, as derived from VLA observations in adjacent sub-bands.
\begin{figure*}
\begin{center}
\includegraphics[angle=90,width=15cm]{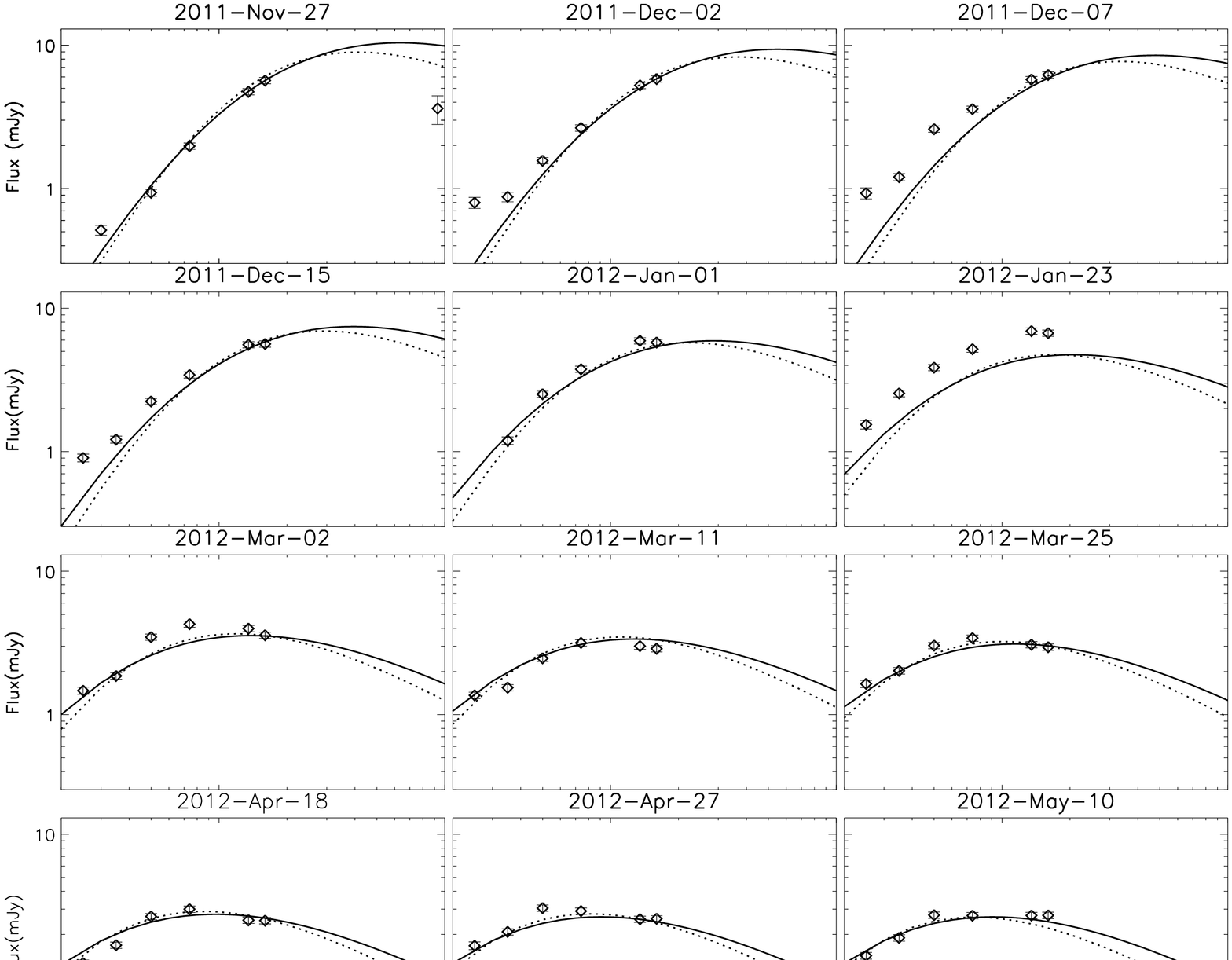}
\vspace{0.5cm}
\caption{Comparison of the observed radio spectra of PTF11qcj at various epochs, and the synchrotron self-absorption model predictions (solid and dotted lines; see text for the models description). We note that the models break down during our latest spectra of this source, when a radio re-brightening is evident. \label{spettri}}
\end{center}
\end{figure*}
The observed spectral indices at early times do not exceed the limiting value of $2.5$ expected in the case of synchrotron self-absorption. In what follows, we thus assume the simplest ``standard'' synchrotron-self absorption model. However, we stress that alternative scenarios invoking internal free-free absorption, or free-free absorption from a clumpy CSM, may also explain such spectral indices and/or contribute to the attenuation mechanisms during the earlier part of our observations \citep{Chevalier1998,Weiler1990}. 

To estimate the physical parameters of PTF11qcj, we follow the formulation by \citet{Soderberg2003L} for synchrotron self-absorbed emission arising from sub-relativistic SN ejecta expanding in a smooth circumstellar medium with a power-law density distribution. Given the observed variability of PTF\,11qcj, we do not expect smooth hydrodynamic models to provide a perfect fit to its radio observations. Nevertheless, we attempt such modeling to provide a tentative estimate of the bulk properties of PTF\,11qcj and its surrounding environment. Hereafter, we consider all the data points collected with the VLA before 56101\,MJD. On and after this epoch, a radio re-brightening is evident (possibly associated with an enhanced density CSM; Figure \ref{radiolc} and \ref{spettri}). Since late-time radio observations of PTF\,11qcj aimed at confirming and tracking this phase of enhanced emission are on-going at the time of writing, here we concentrate on the interpretation of the earlier epochs. 

Within the synchrotron self-absorbed model, at any time $t$ the observed synchrotron emission originates from a shell of shock-accelerated electrons, with radius $r$ and and thickness $r/\eta$. The shell expands spherically while following a self-similar evolution in the interaction with a smooth CSM. The electrons are accelerated into a power-law energy distribution $N(\gamma)\propto \gamma^{-p}$, with $\gamma \gtrsim \gamma_m$, and carry a fraction $\epsilon_e$ of the energy density of the ejecta. A fraction $\epsilon_B$ of such energy goes into magnetic fields. The temporal evolution of the radius ($r$), magnetic field ($B$), minimum Lorentz factor ($\gamma_m$), and of the electrons-to-magnetic field energy ratio ($\epsilon_e/\epsilon_B$), is parametrized as follows \citep{Soderberg2003L,Soderberg2003bg}:
\begin{eqnarray}
\label{scaling1}r= r_0 \left(\frac{t-t_e}{t_0}\right)^{\alpha_r}~~&~~B=B_0 \left(\frac{t-t_{e}}{t_0}\right)^{\alpha_B}~~\\ 
\label{scaling2}\gamma_m=\gamma_{m,0}\left(\frac{t-t_{e}}{t_0}\right)^{\alpha_{\gamma}}~~&~~\frac{\epsilon_e}{\epsilon_B},=\mathfrak{F}_0 \left(\frac{t-t_{e}}{t_0}\right)^{\alpha_{\mathfrak{F}}},
\end{eqnarray}
where $t_0$ is an arbitrary reference time that we set to day 10 since explosion, and $t_e$ is the explosion time of the SN. In the above equations, $\alpha_r=(n-3)/(n-s)$ \citep{Chevalier1982,Chevalier1996}, where n characterizes the density profile of the outer SN ejecta ($\rho_{\rm SN} \propto (r/t)^{-n}$), and $s$ characterizes the density profile of the radiating electrons within the shocked CSM ($n_e \propto r^{-s}$). 

In the ``standard'' scenario \citep{Chevalier1996}, the magnetic energy density ($U_B\propto B^2$) and the relativistic electron energy density ($U_e\propto n_e \gamma_m$) are assumed to be a fixed fraction (i.e., $\alpha_{\mathfrak{F}}=0$) of the total post-shock energy density ($U\propto n_e \left<v\right>^{2}$). With these assumptions \citep{Soderberg2003L,Soderberg2003bg},
\begin{equation}
U_e\propto  U~~\Rightarrow \alpha_{\gamma}=2(\alpha_r-1),\label{ue}
\end{equation}
and 
\begin{equation}
U_B \propto U~~\Rightarrow \alpha_B=\frac{(2-s)}{2}\alpha_r-1\label{ub},
\end{equation} 

With the above parametrization, the flux density from the uniform shell of radiating electrons reads \citep{Soderberg2003L,Soderberg2003bg}:
\begin{eqnarray}
\nonumber f_{\nu}=C_f\left(\frac{t-t_e}{t_0}\right)^{(4\alpha_r-\alpha_B)/2}(1-\exp(-\tau_{\nu}^\xi(t)))^{1/\xi}\left(\frac{\nu}{\rm 1\,GHz}\right)^{5/2}\times\\~\times F_3(x)F^{-1}_2(x)\,{\rm mJy}~~
\label{flussoteoradio}
\end{eqnarray}
where $x=2/3(\nu/\nu_m)$, and
\begin{equation}
\nu_m=\gamma^{2}_m \frac{e B}{2\pi m_e c}=\nu_{m,0}\left(\frac{t-t_e}{t_0}\right)^{2\alpha_{\gamma}+\alpha_B}
\label{num}
\end{equation}
 is the characteristic synchrotron frequency of electrons with Lorentz factor $\gamma_m$; $F_2$ and $F_3$ are integrals of the modified Bessel function of order 2/3 \citep[see Equation (A11) in][]{Soderberg2003L}; $\xi$ parametrizes the sharpness of the spectral break between optically thick and thin regimes \citep{Soderberg2003L,Soderberg2003bg}, and
\begin{equation}
\tau_{\nu}(t)=C_{\tau}\left(\frac{t-t_{e}}{t_0}\right)^{(p-2)\alpha_{\gamma}+(3+p/2)\alpha_B+\alpha_r+\alpha_{\mathfrak{F}}}\left(\frac{\nu}{\rm 1\,GHz}\right)^{-(p+4)/2}F_2(x)
\label{tau}
\end{equation}
is the optical depth \citep{Soderberg2003L,Soderberg2003bg}.

As evident from Equations (\ref{flussoteoradio})-(\ref{tau}), the observed spectral and temporal evolution of the radio emission depends on the parameters $(C_f,C_{\tau},\nu_{m,0},t_e,p,\xi,\alpha_r,s)$. Note that the \textit{three} parameters $C_f$, $C_{\tau}$, and $\nu_{m,0}$ are functions of the \textit{four} normalization constants $r_0$, $B_0$, $\gamma_0$, $\mathfrak{F}_0$\footnote{We refer the reader to Equations (6)-(8) in \citet{Soderberg2003L} for the expression of  $C_f$, $C_{\tau}$, $\nu_{m,0}$, as functions of $r_0$, $B_0$, $\gamma_0$, $\mathfrak{F}_0$.} (and of the source distance $d_L$). Thus, while $C_f$, $C_{\tau}$, and $\nu_{m,0}$ can be determined by comparison with the data, another constraint is needed to derive $r_0$, $B_0$, $\gamma_{m,0}$, $\mathfrak{F}_0$ from the fitted values of $C_f$, $C_{\tau}$, and $\nu_{m,0}$. Following common practice in radio SNe studies, we set $s=2$ (as expected for a wind density profile), $p\approx 3$, and $\nu_0\approx 1$\,GHz. This reduces the number of free parameters to five, $(C_f,C_{\tau},t_e,\xi,\alpha_r)$, or four when the explosion date $t_e$ is known.

In the case of PTF11qcj, we first set $t_e \approx 55842$\,MJD (as suggested by our optical observations; see Sections \ref{spectralproperties} and \ref{Opticallightcurve}). Using a $\chi^2$ minimization procedure and the simplified four-parameter model described above, we obtain the results reported in Table \ref{radiomodeltab} (Model 1). The model predictions are shown in Figures \ref{radiolc}-\ref{spettri} (dotted lines). From our radio best fit results and by Equations (\ref{scaling1}-\ref{scaling2}) here and (6-8) in \citet{Soderberg2003L}, we derive the evolution of $r$, $B$, and $\gamma_m$ (Table \ref{radiomodeltab}). The radial evolution is within the expected range of $0.67\lesssim \alpha_r\lesssim 1.0$ \citep{Chevalier1996,Chevalier1998}, and it implies an ejecta velocity of $dr/dt\approx 0.34c$ at day 10. A velocity of $\approx 0.3\,c$ at day 10, is somewhat higher than the typical range of $(0.1-0.25)\,c$ for normal Ib/c SNe \citep[e.g.,][]{Berger2003}, but significantly slower than SN\,1998bw, for which the bulk Lorentz factor was estimated to be $\Gamma\sim2$ on a similar timescale \citep{Kulkarni1998,Li1999}. We also note that the implied velocity of $\approx 0.28\,c$ at the time of our first spectrum of PTF11qcj, is consistent with the lower-limit estimate derived from the Ca triplet ($\approx 22,000$\,km\,s$^{-1}$), while the implied size at day 100 ($r\approx 7\times 10^{16}$\,cm) is consistent with the lower limit ($r\gtrsim 3\times 10^{16}$\,cm at day 100) derived in Equation (\ref{equi}).

The number density of the emitting electrons, progenitor mass loss rate, and total energy of the ejecta, then follow from the relations \citep{Soderberg2003L,Soderberg2003bg}:
\begin{equation}
n_e=\frac{p-2}{p-1}\frac{B^2_0}{8\pi}\frac{\mathfrak{F}_0}{m_ec^2\gamma_{m,0}}\left(\frac{t-t_{e}}{10\,\rm d}\right)^{2\alpha_B-\alpha_{\gamma}}=n_{e,0}\left(\frac{t-t_{e}}{10\,\rm d}\right)^{\alpha_{n_e}}\label{ennee}
\end{equation}
\begin{equation}
\dot{M}=\frac{8\pi}{\eta}n_{e,0}m_pr_0^2{\rm v}_w\left(\frac{t-t_{e}}{10\,\rm d}\right)^{2\alpha_B-\alpha_{\gamma}+2\alpha_r}=\dot{M}_0\left(\frac{t-t_{e}}{10\,\rm d}\right)^{\alpha_{\dot{M}}}\label{mdot}
\end{equation}
\begin{equation}
E_{K}=\frac{4\pi \mathfrak{F}_0}{\eta}r^3_0\frac{B^2_0}{8\pi\epsilon_e}\left(\frac{t-t_{e}}{10\,\rm d}\right)^{2\alpha_B+3\alpha_{r}}=E_{K,0}\left(\frac{t-t_{e}}{10\,\rm d}\right)^{\alpha_{E_k}}\label{ene}
\end{equation}
for a nucleon-to-electron density ratio of 2 (as appropriate for W-R winds). The values derived for these quantities within Model 1 are also reported in Table \ref{radiomodeltab}. 

\begin{table*}
\begin{center}
\caption{Radio model parameters for a standard synchrotron self-absorbed model with $s=2$ and $\nu_{m,0}=1$\,GHz. All quantities marked with a $_0$ are normalized to $t-t_e=10$\,d. The first five rows give the values of the fitted parameters. The other rows give the values of physical quantities derived by Equations (\ref{scaling1}-\ref{scaling2}), (\ref{ue}-\ref{ub}), (\ref{ennee}-\ref{ene}), and Equations (6-8) in \citet{Soderberg2003L}.\label{radiomodeltab}}
\begin{tabular}{ccc}
 &  Model 1 (fixed $t_e$) & Model 2 (free $t_e$)\\
\hline
$C_f$ & $\approx 7.1\times10^{-4}$ & $\approx 1.9\times10^{-3}$\\
$C_{\tau}$ & $\approx 1.8\times10^{7}$ & $\approx 9.3\times10^6$\\
$\xi$ & $\approx 0.23$ & $\approx 0.17$\\
$\alpha_r$ & $\approx 0.80$ & $\approx 0.80$\\
$t_e$ (MJD) & 55842 & $\approx 55857$ \\
\hline
$r_0$ (cm) & $\approx1.1\times10^{16}\mathfrak{F}_0^{-1/17}(\eta/10)^{1/17}$& $\approx1.7\times10^{16}\mathfrak{F}_0^{-1/17}(\eta/10)^{1/17}$\\
$B_0$ (G) & $\approx6.7\,\mathfrak{F}_0^{-1/17}(\eta/10)^{4/17}\,$ & $\approx5.1\,\mathfrak{F}_0^{-1/17}(\eta/10)^{4/17}\,$\\
$\alpha_B$ & -1 & -1\\
$\gamma_{m,0}$ & $\approx 7.3\,\mathfrak{F}_0^{2/17}(\eta/10)^{-2/17}$ &$\approx 8.3\,\mathfrak{F}_0^{2/17}(\eta/10)^{-2/17}$\\
$\alpha_{\gamma}$ & -0.40 & -0.40\\
$n_{e,0}$ (cm$^{-3}$) & $\approx1.4\times10^{5}\mathfrak{F}_0^{13/17}(\eta/10)^{10/17}$ & $\approx7.6\times10^{4}\mathfrak{F}_0^{13/17}(\eta/10)^{10/17}$\\
$\alpha_{n_{e}}$ &  $\approx -1.6$ & $\approx -1.6$\\
$\dot{M}_0$ ($M_{\odot}{\rm yr}^{-1}$) & $\approx1.2\times10^{-4}\mathfrak{F}_0^{11/17}(\eta/10)^{-5/17}({\rm v}_w/10^3\,{\rm km\,s^{-1}})$& $\approx1.4\times10^{-4}\mathfrak{F}_0^{11/17}(\eta/10)^{-5/17}({\rm v}_w/10^3\,{\rm km\,s^{-1}})$\\
$\alpha_{\dot{M}}$ & 0 & 0 \\ 
$E_{K,0}$ (erg) & $\approx 9.3\times10^{48}(\epsilon_{e}/0.33)^{-1}\mathfrak{F}_0^{12/17}(\eta/10)^{-6/17}$ &$\approx 1.9\times10^{49}(\epsilon_{e}/0.33)^{-1}\mathfrak{F}_0^{12/17}(\eta/10)^{-6/17}$\\
$\alpha_{E_K}$ & 0.40 & 0.40\\
\hline
\end{tabular}
\end{center}
\end{table*}

In terms of total energy in the radio emitting material, PTF11qcj ranks as one of the most energetic Ib/c SNe, comparable to the GRB-associated SN\,1998bw \citep{Kulkarni1998}, and similar to other radio SNe that show evidence for CSM density variations \citep[e.g., SN\,2003bg;][]{Soderberg2003bg}. The mass-loss rate estimates for GRBs and engine-driven SNe such as 1998bw are typically $(\dot{M}/10^{-5} M_{\odot}{\rm yr}^{-1})\times ({\rm v_w/10^3\,km\,s^{-1}})^{-1}\approx 0.1-1$ \citep{Panaitescu2002,Yost2003,Chevalier2004}. PTF11qcj lies at the higher end of the range observed for local W-R stars, $\dot{M}\approx (0.6-9.5)\times10^{-5}M_{\odot}\,{\rm yr}^{-1}$ \citep{Cappa2004}, and is similar to \textit{normal} type Ib/c SNe with prominent variations in the radio light curves \citep[such as SNe 2004cc and 2004gq;][]{Soderberg2003bg,Wellons2012}. 

 As underlined before, we do not expect the simplified model used in our analysis to provide a perfect fit to (nor to represent a complete physical interpretation of the) PTF11qcj radio emission. Indeed, our high $\chi^2/$d.o.f=$1790/90$ is similar to what obtained in other analyses of complex radio SN light curves \citep[e.g.,][]{Soderberg2003bg,Krauss2012}. So the results of our fit should be regarded as a tentative insight into the properties of the PTF11qcj fastest ejecta and CSM. 

The match between Model 1 and data is most problematic during the earlier observations at the lowest frequencies ($2.5-3.5$\,GHz): we caution, however, that at such low frequencies the data are most affected by RFI, so the uncertainties on the measured fluxes may be underestimated.  

We note that it is unlikely that the high value of the $\chi^2$ derived for the radio model is entirely dominated by interstellar scattering and scintillation (ISS). Indeed, considering the data collected at $t\leq 56057$\,MJD (i.e., in between $\approx 38-215$\,d since 55842\,MJD), we derive a modulation index with respect to our best fit model flux predictions ($f_{\nu,pred}$),
\begin{equation}
m_{obs}(\nu)=\frac{\sqrt{\left<(f_{\nu,obs}-f_{\nu,pred})^{2}\right>-\left<\sigma^{2}_{\nu,obs}\right>}}{\left<f_{\nu,obs}\right>}
\end{equation}
(where $f_{\nu}$ are the measured flux densities and $\sigma_{\nu}$ their measured uncertainties), of $m(\nu)\approx 0.1-0.3$ for $\nu=(2.5-16)$\,GHz (with the highest modulation observed at the lowest frequencies). Using the maps provided by \citet{Walker2001} and the PTF11qcj Galactic coordinates ($l=112\deg$, $b=69\deg$), we roughly estimate $\nu_0\approx 5$\,GHz for the ISS transition frequency between the strong and weak scattering regimes (thus, our lowest frequency observations are in the strong scattering regime), and $\Theta_0\approx 5\,\mu$as for the Fresnel angle at the transition frequency. The expected value for the modulation index in the refractive strong scattering reads \citep[e.g.,][]{Walker1998,Walker2001,Kulkarni1998,Cenko2013}:
\begin{eqnarray}
m_{p}(\nu)=\left(\frac{\nu}{\nu_0}\right)^{17/30}\left(\frac{\Theta_r}{\Theta}\right)^{7/6},
\end{eqnarray}
with
\begin{equation}
\Theta_r=\Theta_0\left(\frac{\nu_0}{\nu}\right)^{11/5}
\end{equation}
Using the above Equations, $m_p({\rm 2.5\,GHz})\approx 0.3$ requires $\Theta \approx 45\,\mu$as around day 130, or $r\approx 4\times 10^{16}$\,cm, which is a factor of $\approx 2$ smaller than the radius at day 130 implied by our radio fits. Moreover, flux variations are observed up to the highest frequencies of our observations (16\,GHz), that fall in the weak scattering regime (and so the observed modulations at such frequencies cannot be explained by ISS).

Due to the uncertainties in the explosion date of PTF\,11qcj, we also performed a fit removing the constraint on the explosion time (Model 2; see Table \ref{radiomodeltab} and the solid lines in Figures \ref{radiolc}-\ref{spettri}). In this case the fit drives $t_e$ around the discovery time of PTF\,11qcj, $55857$\,MJD, and $\chi^2_r$/d.o.f.=1717/89. We note, however, that in this last model the velocity of the fastest ejecta is rather high, $dr/dt\approx 0.52c$ at day 10. Assuming PTF11qcj exploded around $55842$\,MJD (see Section \ref{spectralproperties}), the delayed onset of the radio emission may be interpreted as interaction with a shell whose inner radius is located at a distance $r\gtrsim 22,000$\,km\,s$^{-1}\times 15$\,d\,$\approx 3\times10^{15}$\,cm. This is comparable to the distances that may have been reached by material ejected during the pre-SN activity event probed by our P48 pre-discovery images (see Section \ref{LBVburst}), if such material was ejected at speeds v$_w\gtrsim 400$\,km\,s$^{-1}$.

An element common to both the above fit results is the low value of $\xi\approx 0.2$, a parameter that measures the sharpness of the spectral break between optically thick and thin regimes. Indeed, we see that at the higher frequencies the late-time spectral index of PTF11qcj appears to remain $\gtrsim -1$, or show an abrupt increase (this is evident during our latest two VLA observations; Figure \ref{spectralindex} and Figure \ref{spettri}) rather than transition toward a value of $\approx -1$. Recently, \citet{Bjornsson2013} has shown that the broadening observed in the radio spectra and/or light curves of some type Ib/c SNe may be a direct indication of inhomogeneities in the CSM.

The increase in spectral index observed for PTF11qcj at $t\gtrsim 300$\,d (during the re-brightening phase) in the 2.5-13\,GHz range is reminiscent of cases like the normal type Ic SN\,2004cc, for which light curve variations associated with an increase of the spectral peak have been observed \citep{Wellons2012}, or the Ic-BL SN\,2007bg, for which an increase in the flux density accompanied by an increase in spectral index was observed around day 300 and interpreted as an absorption turn-on associated with a sharp CSM density enhancement presumably due to evolution of the progenitor wind \citep{Salas2012}. Indeed, this event seems most similar to PTF11qcj (see also Figure \ref{radio_comparison}). According to our radio modeling of the earlier data, the re-brightening occurs at $r\gtrsim1.7\times10^{17}$\,cm. 

Finally, we note that \citet{Moriya2013} have recently investigated the link between LBV as SN progenitors and the appearance of episodic light curve modulations in the radio light curves of the SN events, showing how these early modulations have the potential to probe the progenitor's mass loss history immediately before the SN explosion. A long-term monitoring campaign of PTF11qcj aimed at tracking the properties of its late-time re-brightening is on-going at the time of writing.
\subsection{X-ray emission modeling}
\begin{figure}
\begin{center}
\includegraphics[width=9cm]{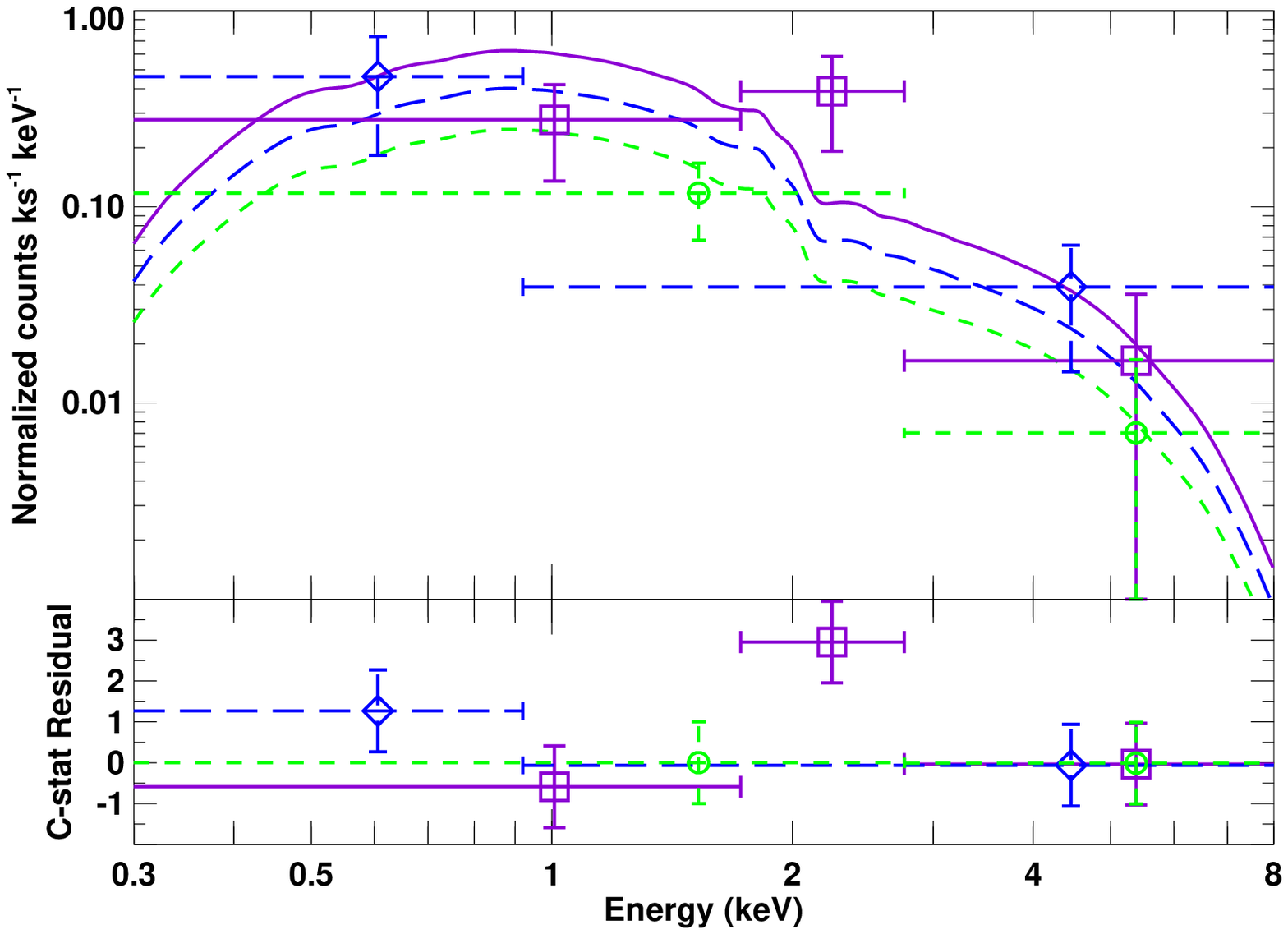}
\caption{\textit{Chandra} X-ray spectral fit to our three observations of
PTF11qcj. Top panel: Best-fit power-law spectra (smooth lines) and
binned data (points with error bars) for the three epochs (first
epoch: purple squares and solid line; second epoch: blue diamonds and
long dashed line; third epoch: green circles and short dashed line). The
multiepoch spectral fit has hydrogen-equivalent absorbing column fixed
at the estimated Galactic column of $N_H = 10^{20}$\,cm$^{-2}$, fluxes
allowed to vary at each epoch, and power-law index fixed across the
three epochs; this fit yields photon index $\Gamma =
1.70^{+0.81}_{-0.72}$ (90\%-confidence range).  Bottom panel:
C-statistic residuals to the fit; numerical simulations give an
acceptable goodness metric of 30\%, and no obvious defects are
apparent. \label{xfig}}
\end{center}
\end{figure}

We performed a spectral fit of our three observations of PTF11qcj with \textit{Chandra} (Figure \ref{xfig}). We used a power-law model with fixed $N_{H}$, and power-law index $\Gamma$ (where $N_{\rm phot}(E)\propto E^{-\Gamma}$) free to vary but constrained to be the same across the three epochs. Flux normalizations were set free to vary for each epoch. This way, we derive $\Gamma=1.70^{+0.81}_{-0.72}$ ($90\%$ confidence; Cstat/d.o.f.=5.18/3) for $N_{H}=10^{20}$\,cm$^{-2}$ (Galactic column density in the direction of PTF11qcj). We achieve satisfactory X-ray spectral fits for fixed $N_H$ and power-law photon index unchanging across the three epochs, as long as $N_H$ is less than a few $\times 10^{21}$\,cm$^{-2}$. Even if the CSM is not ionized, this upper-limit on $N_H$ is consistent with the mass-loss (and radius at day 100) derived from our radio analysis. Indeed, for a given $\dot{M}$ and $r$, the implied $N_H$ (from $r$ to infinity) reads \citep[e.g.,][]{Ofek2013}:
\begin{eqnarray}
\nonumber N_H\approx 10^{20} \left(\frac{\dot{M}}{1.5\times 10^{-4} M_{\odot}\,\rm yr^{-1}}\right)\left(\frac{v_w}{10^3{\rm\,km\,s^{-1}}}\right)^{-1}\\\times \left(\frac{r}{7\times10^{16}\rm\,cm}\right)^{-1}\,{\rm cm^{-2}}~~.
\end{eqnarray}
 
From the above spectral analysis, we derive the corresponding unabsorbed X-ray fluxes for the three epochs (see the last column in Table \ref{X}). If we conservatively assume that all the flux measured during the last of our \textit{Chandra} observations is entirely due to the contribution of the host galaxy, and subtract such flux from the flux measured during the first epoch, then we get that the X-ray emission from PTF11qcj is $\gtrsim (4.5\pm2.9)\times10^{-15}$\,erg\,cm$^{-2}$\,s$^{-1}$, which corresponds to an X-ray luminosity $\gtrsim (8\pm5)\times10^{39}$\,erg\,s$^{-1}$ (0.3-8\,keV; cps-to-flux conversion computed assuming the best fit $\Gamma\approx 1.7$, and an $N_{H}\approx 10^{20}$\,cm$^{-2}$). 

Hereafter, we consider three main mechanism to explain PTF11qcj X-ray observations \citep{Chevalier2006}: synchrotron emission; thermal (free-free) bremsstrahlung emission from material in the circumstellar shock or in the ejecta reverse shock; and inverse Compton scattering of photospheric emission photons by relativistic electrons. 

The radio-to-X-ray spectral index estimated comparing the 16\,GHz flux of PTF11qcj around the time of our first \textit{Chandra} observation ($F_{\rm 16\,GHz}\approx 6.5$\,mJy) with the X-ray one ($F_{\rm X}\gtrsim 4.5\times10^{-15}\,\rm{erg\,cm^{-2}\,s^{-1}}/(2.4\times10^{17}\rm{\,Hz})\approx 2\times10^{-6}$\,mJy), is $\Gamma-1\approx0.9$. The last is in agreement with both the X-ray spectral index derived from the analysis of our \textit{Chandra} data, and the radio synchrotron self-absorbed model discussed in the previous section. Indeed, within the last model, the optically thin portion of the synchrotron spectrum is predicted to have a photon index of $\Gamma-1\approx (p-1)/2\approx 1$. 

Based on the above considerations, synchrotron emission would seem to be the most straightforward interpretation of the PTF11qcj X-ray observations. A complication (not unique to the case of PTF11qcj) arises from the fact that the synchrotron cooling break is expected to steepen the spectral index between the radio and X-ray bands. For PTF11qcj, this would cause the X-ray flux extrapolated from our radio data to fall below the one measured by \textit{Chandra}. However, \citet{Ellison2000} have suggested that a cosmic-ray dominated shock can flatten the synchrotron spectrum at high frequencies, and \citet{Chevalier2006} have invoked such a flattening to explain the X-ray emission of SN\,1994I in the context of synchrotron emission.

Within the thermal emission hypothesis, X-rays are produced while the forward shock plows into the CSM and/or by the reverse shock heating the ejecta. Thus, the X-ray luminosity depends on the density of the emitting material, which then cools by free-free emission. For an $r^{-2}$ wind density profile, the free-free luminosity can be estimated as \citep[e.g.,][]{Chevalier2001,Sutaria2003,Soderberg2003L,Soderberg2003bg,Ofek2013b}:
\begin{eqnarray}
\nonumber L_X\approx \frac{C}{690} \times 3\times 10^{39}\left(\frac{\dot{M}}{1.5\times 10^{-4} M_{\odot}{\rm yr^{-1}}}\right)^{2}\left(\frac{{\rm v}_w}{10^3\,{\rm km\,s^{-1}}}\right)^{-2}\\\times \left(\frac{r}{5\times 10^{15}\,\rm cm}\right)^{-1}\,{\rm erg\,s^{-1}},~~~~
\end{eqnarray}
where $C=4(1+(n-3)(n-4)^{2}/4)$ accounts for both the reverse and forward shock contribution\footnote{The factor of 4 accounts for the fact that the density behind the forward shock ($\rho_{FS}$) is 4 times larger than the CSM one; moreover, the expression for C takes into account the fact that $\rho_{RS}=\rho_{FS}(n-4)^{2}(n-3)/4$ \citep{Chevalier2001}.}. As evident from the above Equation, for typical values of $n=5-12.5$ \citep[i.e., $C\approx 6-690$ or $\alpha_r\approx 0.67-0.9$;][]{Chevalier1982} and $\dot{M}\lesssim 1.5\times 10^{-4}M_{\odot}{\rm yr^{-1}}({\rm v}_w/10^3\,{\rm km\,s^{-1}})$ \citep[as typically found in Ib/c SNe;][]{Wellons2012}, $L_X\gtrsim (8\pm5)\times10^{39}$\,erg\,s$^{-1}$ at 100\,d requires r (100\,d)$\lesssim5\times 10^{15}$\,cm. This is smaller than the r\,(100\,d)$\gtrsim 3\times 10^{16}$\,cm derived from Equation \ref{equi}, and r\,(100\,d)$\approx 7\times 10^{16}$\,cm derived from the radio modeling (see Table \ref{radiomodeltab}).

In the inverse Compton scenario, the X-ray luminosity can be estimated as follows \citep[e.g.,][]{Bjornsson2004,Soderberg2003L,Soderberg2003bg}:
\begin{equation}
\frac{L_{X}}{L_{\rm radio}}\approx \frac{U_{\rm ph}}{U_B}
\end{equation}
where:
\begin{equation}
U_B=\frac{B^{2}}{8\pi}
\end{equation}
and \citep{Soderberg2003bg}:
\begin{equation}
U_{\rm ph}\approx0.4 \left(\frac{L_{\rm bol}}{10^{42}\rm erg\,s^{-1}}\right)\left(\frac{t}{\rm 1d}\frac{\rm v}{c}\right)^{-2}\,{\rm erg\,cm^{-3}}.
\end{equation}
For PTF11qcj, using our late-time P48 photometry and applying a tentative bolometric correction of $\approx -0.5$ \citep[for a SN temperature of $\approx 10^4$\,K;][]{Corsi2012}, we estimate $L_{\rm bol}\approx 10^{8}L_{\odot}$ during the first \textit{Chandra} epoch. Also, at this time, we estimate ${\rm v}=dr/dt\approx 0.2\,c$ and $B\approx 0.7$\,G based on the radio model discussed in Section \ref{radiomodel}. Thus, we get $U_{\rm ph}\approx 4\times10^{-4}$\,erg\,cm$^{-3}$ and $U_B\approx 2\times10^{-2}$\,erg\,cm$^{-3}$, or $U_{\rm ph}/U_B\approx 0.02$. Since $L_{\rm radio}\approx 2\times 10^{39}$\,erg\,s$^{-1}$ (based on the 16\,GHz flux of $\approx 6.5$\,mJy during the first \textit{Chandra} observation), we derive $L_{X}\approx 4\times 10^{37}$\,erg\,s$^{-1}$, which underestimates the observed X-ray luminosity by a factor of $\gtrsim 100$.

 In conclusion, synchrotron emission seems to be the easiest explanation for the X-ray counterpart to PTF11qcj but, as discussed before, a mechanism flattening the spectrum at high frequencies (such as a cosmic-ray dominated shock) would be required. Free-free emission could explain the observed counterpart only for mass-loss rates in excess of the highest values found for SNe of type Ib/c ($\dot{M}\gtrsim 1.5\times10^{-4}\,M_{\odot}\,$yr$^{-1}$); the inverse Compton scenario seems hard to reconcile with the observed X-ray counterpart to PTF\,11qcj, unless the magnetic field energy is much lower than what estimated from the radio fit \citep[which assumes equipartition; see also][]{Horesh2012}.
\subsection{Search for $\gamma$-rays}
Given the radio and X-ray detections of PTF11qcj, we have searched the \textit{Fermi} GRB catalog for GRBs with explosion date in between the time of our first detection of PTF11qcj, and a month earlier. No GRBs were found within such time frame, with positions compatible with PTF11qcj. The non-relativistic speeds suggested by the simplest radio model (Section \ref{radiomodel}) are in agreement with the lack of a GRB detection.

From a tentative comparison with the light curves of off-axis low-luminosity GRBs expanding in a constant density environment available in the literature \citep{van2011}, we deduce that fitting the PTF11qcj radio and X-ray emission within the \textit{simplest} off-axis GRB scenarios, may be difficult to achieve. Moreover, despite there is some amount of controversy in the literature concerning whether the contribution from a GRB counter jet could cause a late-time light curve bump \citep[e.g.,][]{Li2004,Wang2009,van2011}, the late-time radio re-brightening observed in PTF11qcj is a feature that challenges the simplest off-axis models. 
\subsection{IR echo}
\begin{figure*}
\begin{center}
\includegraphics[width=12.cm]{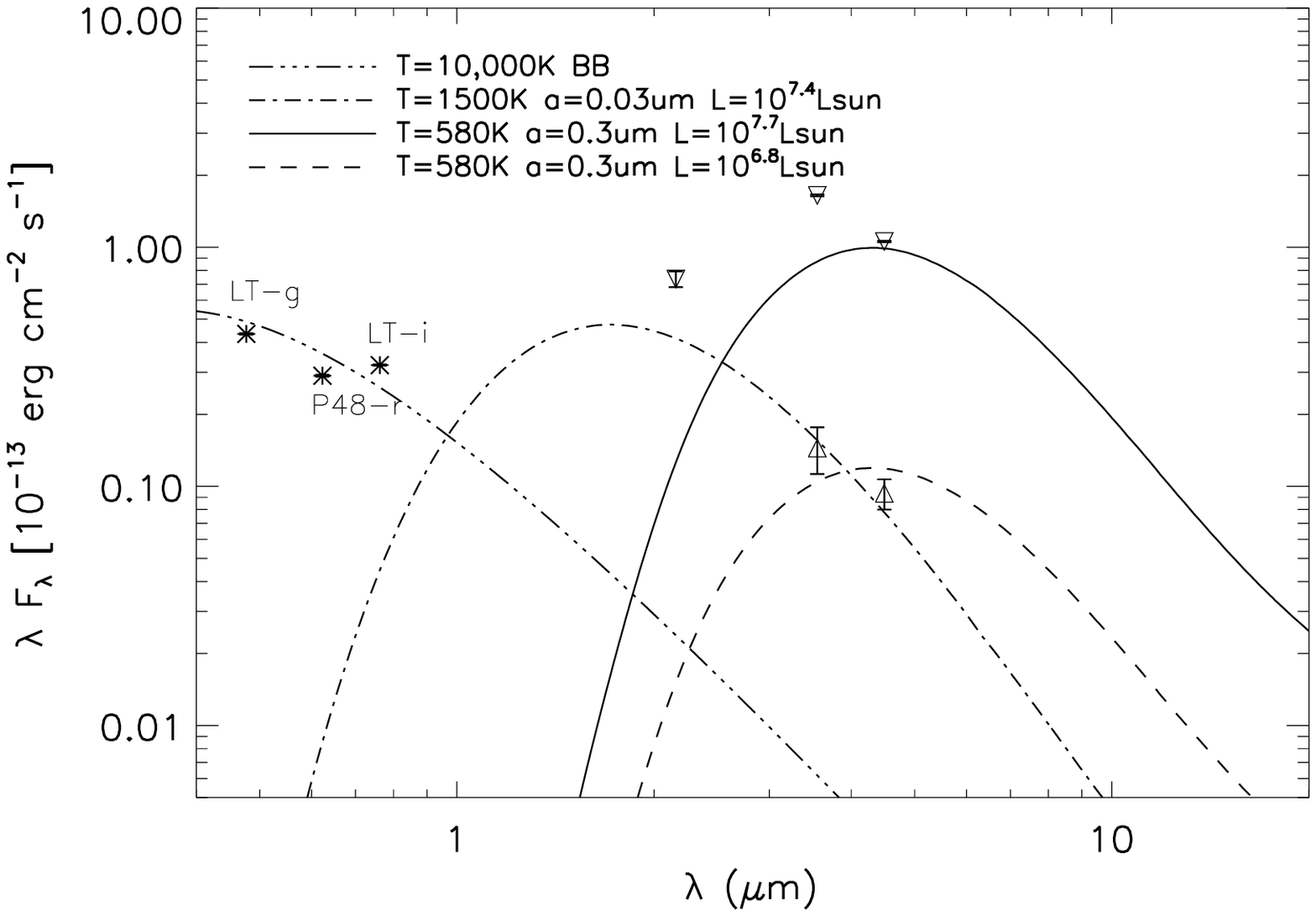}
\caption{\textit{Spitzer} and P200 data points (downward pointing triangles) collected on 2012 March plotted as upper limits (due to the contribution from the host galaxy emission, see Section \ref{IRobservations}), and P48 and LT $gri$ photometric data points around the same epoch. Plotted as lower-limits (upward pointing triangles) are the data obtained by subtracting from the flux measured by \textit{Spitzer} on 2012 March, the one measured on 2012 June (see Figure~\ref{spitzer}). Data are compared with the spectrum that would be expected for dust grains of size $0.03\mu$\,m, temperature $T_{\rm echo}\approx 1600$\,K, and dust mass $M_{d}\approx 10^{-5}$\,M$_{\odot}$ (or $10^{7.4}\,L_{\odot}$; dash-dotted line); and with the spectra expected for dust grain of size $0.3\mu$m, temperature $T_{\rm echo}\approx 580$\,K, and dust mass in between $M_{d}\approx 2\times 10^{-4}$\,M$_{\odot}$ (or $10^{6.8}\,L_{\odot}$; dashed line) and $M_{d}\approx 10^{-3}$\,M$_{\odot}$ (or $10^{7.7}\,L_{\odot}$; solid line). For reference, we also plot a 10,000\,K black body spectrum normalized so as to match the P48 and LT data, showing that the contribution from the SN light to the IR is negligible compared to our IR lower-limits. \label{BB}}
\end{center}
\end{figure*}

We detected an IR counterpart to PTF11qcj in both the \textit{Spitzer} (Figure \ref{spitzer}) and P200 (Figure \ref{scoperta}) images. The extrapolation from the visible light spectral energy distribution suggests that a late-time IR excess may be associated with PTF11qcj (Figure \ref{BB}). A continuum IR emission can be expected from hot dust - either newly formed hot dust, or pre-existing dust in the CSM that is heated by the SN. Several heating mechanisms are possible. Pre-existing dust, for example, may be collisionally heated by hot shocked gas, or radiatively heated by either the peak SN luminosity or the late-time optical emission from circumstellar interaction \citep[e.g.,][]{Draine1979,Draine1981,Dwek1983,Dwek1985}. 

Detailed modeling of the dust components and heating mechanisms is hard to accomplish with only upper limits. Nonetheless, we can get some tentative insight into the properties of the dust using our \textit{Spitzer} and P200 observations. The following analysis should be considered as an order of magnitude estimate of the dust properties around PTF11qcj, rather than an accurate measurement. 

The simplest scenario usually invoked to explain IR echos observed in type II SNe is that of radiative heating of pre-existing dust by the SN optical emission. The SN peak luminosity is expected to vaporize the dust out to a vaporization radius $r_{\rm vap}$, and heat the inside of this shell to a temperature roughly equal to the vaporization temperature, $T_{\rm vap}\approx 2000$\,K for graphite (hereafter, we limit our discussion to graphite grains only since our data do not allow us to distinguish between e.g. graphite and silicate). Alternatively, the dust may be formed at a radius $r_{\rm echo}$ larger than the vaporization radius, and heated to a temperature $T_{\rm echo}\lesssim T_{\rm vap}$. 

For simple dust populations composed entirely of graphite with a single grains size, we can estimate the echo (vaporization) radius ($r_{\rm echo / vap}$) of a dust shell at $T_{\rm echo}$\ ($T_{\rm vap}\approx 2000$\,K), using the estimated peak bolometric luminosity of PTF11qcj ($L_{\rm bol, p}$), and the fact that the equilibrium dust temperature is set by balancing the energy absorbed by a spherical dust grain of radius $a$ \citep{Fox2010}:
\begin{eqnarray}
\nonumber L_{\rm abs}(a)=4\pi r^{2}_{\rm SN}\frac{\pi a^{2}}{4\pi r^{2}_{\rm echo/vap}}\int \pi B_{\nu}(T_{\rm SN}) Q_{\nu}(a) d\nu=\\=\frac{L_{\rm bol,p}}{\sigma T^{4}_{\rm SN}}\frac{ a^{2}}{4 r^{2}_{\rm echo/vap}}\int \pi B_{\nu}(T_{\rm SN}) Q_{\nu}(a) d\nu,
\end{eqnarray}
and the energy emitted by the dust:
\begin{equation}
L_{\rm rad}(a)=4\pi a^{2} \int \pi B_{\nu}(T_{\rm echo/vap}) Q_{\nu}(a) d\nu.
\label{lumgrain}
\end{equation}
In the above Equations we have assumed wavelength dependent emissivities/absorption $Q_{\nu}(a)$ \citep{Draine1984,Laor1993}; $B_{\nu}(T)$ is the Planck black-body function; $\sigma$ is the Stefan-Boltzmann constant; $T_{\rm SN}\approx 10^4$\,K is the effective SN temperature; and $r_{\rm SN}$ the effective SN emitting radius. The equality condition yields \citep{Fox2010}:
\begin{equation}
r_{\rm echo/vap}(a)=\sqrt{\frac{3}{64}\frac{L_{\rm bol,p}}{(\pi \rho_d a \sigma T_{\rm SN}^4)} \frac{\int B_{\nu}(T_{\rm SN})Q_{\nu}(a)d\nu}{\int B_{\nu}(T_{\rm echo/vap})k_{\nu}(a)d\nu}}
\label{rev}
\end{equation} 
where $k(\nu)=Q_{\nu}(a)/(4/3\rho_d a)$ is the mass absorption coefficient, and $\rho_d=2.25$\,g\,cm$^{-3}$ \citep{Smith2008} is the assumed dust density.

To estimate PTF11qcj peak bolometric luminosity we use the observed P48 peak magnitude corrected for Galactic and local host galaxy extinction, and apply a bolometric correction of $-0.5$\,mag \citep[for a SN temperature of $\approx 10^4$\,K;][]{Corsi2012}. This way we get $L_{\rm bol,p}\approx 10^{9.2}L_{\odot}$ (but note that our observations do not allow us to precisely constrain the optical peak). Thus, from Equation (\ref{rev}):
\begin{eqnarray}
r_{\rm vap}\approx (1-5)\times 10^{16}\,{\rm cm}.
\end{eqnarray}
The above range is calculated for dust grains of sizes $(0.03-0.3)\,\mu$m, and the smaller $r_{\rm vap}$ are for the larger dust grains (note that in Equation (\ref{rev}) $r_{\rm vap}\sim a^{-1}$). 

In the radiative heating scenario, the shock radius ($r_s$) sets a lower limit for the dust radius (since any dust within this radius would be independently heated or destroyed by the forward shock). At the time of our first \textit{Spitzer} observation:
\begin{equation}
r_s \approx (0.3-1)\times10^{17}\,{\rm\,cm}
\label{shockradius}
\end{equation}
The smallest value above is calculated assuming an average speed of $22,000$\,km\,s$^{-1}$ as indicated by the optical spectra of PTF11qcj (Section \ref{spectralproperties}), while the largest value is derived from the radio model (radio emission probes the fastest ejecta; see Section \ref{radiomodel}). Moreover, if the dust shell is located at or beyond the vaporization radius, light travel time effects cause the thermal radiation from the dust grains to reach the observer over an extended period of time (thus the term ``IR echo''). So the duration of the IR echo sets a scale for the dust radius \citep{Fox2010,Fox2011}:
\begin{equation}
c \times \Delta t_{\rm IR} \approx 2 \times r_{\rm IR}.
\label{echorad}
\end{equation}
Using the fact that in the case of PTF11qcj $\Delta t_{\rm IR}\gtrsim 100-200$\,d, we deduce $r_{\rm IR}\gtrsim (1-3)\times10^{17}$\,cm (which is indeed larger than, or at the least comparable to, $r_s$ - this assures self-consistency of the IR echo scenario). 

Since $r_{\rm vap} <r_{\rm IR}$, we conclude that the dust should be at $T_{\rm echo} < T_{\rm vap}$ in order for the IR echo scenario to be self-consistent. Using again Equation (\ref{rev}), and setting the condition $r_{\rm echo} \approx r_{\rm IR}\gtrsim (1-3)\times10^{17}$\,cm, we derive the maximum temperatures that the dust can reach when heated by PTF11qcj optical peak:
\begin{equation}
T_{\rm echo} \lesssim 580-1500\,{\rm K},
\end{equation} 
where the smallest temperature is for dust of size $0.3\,\mu$m located at a dust echo radius of $r_{\rm echo} \approx 3\times10^{17}$\,cm, while the largest temperature is for dust grains of $0.03\,\mu$m located at $r_{\rm echo} \approx 10^{17}$\,cm. We note that temperatures in the range of 500-770\,K are typically found in modeling type IIn SNe emission with graphite dust grains of sizes $a\approx 0.01-0.3\,\mu$m \citep{Fox2011}.

For optically thin dust with mass $M_d$, located at a distance $d_L$ from the observer, and thermally emitting at a single equilibrium temperature $T_{\rm echo}$ (note that the IR emission of SNe such as SN\,2006jd and SN\,2005ip was best fit by a multi-component dust model as opposed to a single component one; here we limit our discussion to a single component - a similar approach has been used for other SNe with limited data-sets; see e.g. \citet{Fox2011}), the total flux reads \citep[see e.g.][]{Fox2010}:
\begin{equation}
F_{\nu}=\frac{M_d}{4/3\pi \rho_d a^{3}}\frac{L_{\nu}(a)}{4\pi d^2_L}=\frac{M_d B_{\nu}(T_{\rm echo})k_{\nu}(a)}{d^2_L},
\label{massa}
\end{equation}
where $L_{\nu}(a)$ is defined in Equation (\ref{lumgrain}). As shown in Figure \ref{BB}, for dust grains of sizes $0.03-0.3\,\mu$m we tentatively estimate dust masses in the range $M_d \sim 10^{-5}-10^{-3}$\,M$_{\odot}$. 
For comparison, dust masses in the range $M_{d}\approx 2\times10^{-5}-5\times 10^{-2}\,M_{\odot}$ are typically invoked for type IIn SNe with IR emission \citep{Fox2011}. 

The estimates derived in this section are subject to the large uncertainties related to the limitation of our IR data-set. However, a light echo scenario from pre-existing dust, presumably emitted by the SN progenitor before the SN itself, seems to offer a consistent explanation.   
\section{Summary}
\label{conclusion}
We have presented panchromatic follow-up observations of SN PTF11qcj, that span an epoch of up to $\approx 1.5$\,yr since optical discovery. While longer-time follow-up at radio wavelengths is still on-going at the time of writing, the radio (VLA and CARMA), IR (\textit{Spitzer} and P200), optical (P48, P60, LT, WHT, and Keck), and X-ray (\textit{Swift} and \textit{Chandra}) data presented here allow us to estimate the mass-loss, energy, and mean expansion velocity of this interacting SN. 
PTF11qcj is as radio-luminous as the GRB-associated SN\,1998bw, and comparably energetic. However, its fastest-moving ejecta are slower, and appear to be expanding in a CSM of density substantially higher than usually observed in engine-driven SNe, but comparable to other Ib/c SNe showing prominent radio light curve variations. 

Our X-ray observations likely suffer from some contamination from PTF11qcj host galaxy. We have conservatively set a lower-limit to PTF11qcj X-ray emission by subtracting the flux measured during the third epoch of our \textit{Chandra} observations to the one observed during the first epoch. All the models we considered to explain PTF11qcj X-ray observations tend to predict an X-ray flux that falls below our lower-limit to PTF11qcj X-ray counterpart. Synchrotron emission may offer a viable explanation of our X-ray measurements if a mechanism flattening the spectrum at high frequencies (such as a cosmic-ray dominated shock) is invoked. Free-free emission could explain the observed counterpart only for mass-loss rates in excess of the highest values found for SNe of type Ib/c ($\dot{M}\gtrsim 1.5\times10^{-4}\,M_{\odot}\,$yr$^{-1}$). 

The IR excess detected at the location of PTF11qcj suggests an IR echo associated with radiative heating of pre-existing dust by the SN optical emission. The order-of-magnitude estimates derived for the values of dust mass and temperature, seem to be compatible with those of type IIn SNe. However, these estimates are subject to the large uncertainties related to the contamination from the SN host galaxy. Observations of the PTF11qcj host galaxy aimed at better estimating the level of contamination to the IR counterpart of PTF11qcj will be the subject of a future paper.

Our optical observations of the PTF11qcj field about 2\,yr before discovery, show tentative evidence for a precursor eruption that could have emitted the material with which the SN is observed to interact. However, we can exclude a precursor as bright as the one observed 2\,yr before the Ibn SN\,2006jc. Given the relatively large amount of mass-loss rate deduced from the radio modeling ($\dot{M}\sim 10^{-4}\,M_{\odot}$yr$^{-1}$), it may be possible that some of this mass was lost in an eruptive process. Indeed, our radio observations presumably require a non-smooth density profile that can be explained by periods of higher-than average mass-loss rates (via eruptions) from the PTF11qcj progenitor.

Overall, PTF11qcj seems to be consistent with the explosion of a massive W-R star. Future data from our long-term monitoring campaign of this source will allow us to further constrain the properties of this rare SN.
\acknowledgments
PTF is a collaboration of Caltech, LCOGT, the Weizmann Institute, LBNL,
Oxford, Columbia, IPAC, and Berkeley. Staff and computational resources
were provided by NERSC, supported by the DOE Office of Science. 
HET/LRS are supported by UT/Austin, the Pennsylvania State University,
Stanford, Ludwig-Maximilians-Universitat Munchen,
Georg-August-Universitat Gottingen, and the Instituto de Astronomia
de la Universidad Nacional Autonoma de Mexico. Support for CARMA 
construction was derived from the Gordon and Betty Moore Foundation, 
the Kenneth T. and Eileen L. Norris Foundation, the James S. McDonnell Foundation, the Associates of the California Institute of Technology, the University of Chicago, 
the states of California, Illinois, and Maryland, and the National Science Foundation. 
Ongoing CARMA development and operations are supported by the National 
Science Foundation under a cooperative agreement, and by the CARMA partner universities.
 The K. Jansky Very Large Array is operated by NRAO, 
for the NSF under cooperative agreement by Associated Universities, Inc.  
The W.M. Keck Observatory, is operated as a scientific partnership 
among the California Institute of Technology, the University of 
California and the National Aeronautics and Space Administration. 
The Observatory was made possible by the generous financial support 
of the W.M. Keck Foundation. 
The Liverpool Telescope is operated on the island of La Palma by
 Liverpool John Moores University in the Spanish Observatorio del 
Roque de los Muchachos of the Instituto de Astrofisica de Canarias 
with financial support from the UK Science and Technology Facilities Council.
The William Herschel Telescope is operated on the island of La Palma by 
the Isaac Newton Group in the Spanish Observatorio del Roque de los 
Muchachos of the Instituto de Astrofísica de Canarias. 
A.G. and S.R.K. acknowledge support from the BSF; 
A.G. further acknowledges support from the ISF,
EU/FP7 via an ERC grant, GIF, Minerva, and a Kimmel Award;
E.O.O. is incumbent of the Arye Dissentshik career development 
chair and is grateful to support by
a grant from the Israeli Ministry of Science and
the I-CORE Program of the Planning
and Budgeting Committee and The Israel Science Foundation (grant No 1829/12);
J.S.B. acknowledges support of an NSF-CDI Grant 0941742, 
``Real- time Classification of Massive Time-series Data Streams'';
M.M.K. acknowledges generous support from the
Hubble Fellowship and Carnegie-Princeton Fellowship;
M.S. acknowledges support from the Royal Society;
S.B.C. acknowledges generous ﬁnancial assistance 
from Gary and Cynthia Bengier, the Richard and 
Rhoda Goldman Fund, the Christopher R. Redlich Fund, 
the TABASGO Foundation, and NSF grant AST-1211916.
A.C. would like to thank the VLA staff for their support, and in particular: 
Miriam Krauss for very useful discussions on many aspects of the data reduction procedures; Heidi Medlin
for support with the scheduling of the observations; and Drew Medlin for useful
discussions on the VLA data reduction pipeline. A.C. also thanks E. Nakar for useful discussions. We thank the anonymous Referee for useful comments.
\bibliographystyle{apj}
\bibliography{Cors2711}

\begin{thebibliography}{}

\bibitem[\protect\citeauthoryear{{Arnett}}{{Arnett}}{1982}]{Arnett1982}
{Arnett}, W.~D. 1982, \apj, 253, 785

\bibitem[\protect\citeauthoryear{{Begelman} \& {Sarazin}}{{Begelman} \&
  {Sarazin}}{1986}]{Begelman1986}
{Begelman}, M.~C.,  \& {Sarazin}, C.~L. 1986, \apjl, 302, L59

\bibitem[\protect\citeauthoryear{{Berger} et~al.}{{Berger}
  et~al.}{2003}]{Berger2003}
{Berger}, E., {Kulkarni}, S.~R., {Frail}, D.~A.,  \& {Soderberg}, A.~M. 2003,
  \apj, 599, 408

\bibitem[\protect\citeauthoryear{{Bj{\"o}rnsson}}{{Bj{\"o}rnsson}}{2013}]{Bjor%
nsson2013}
{Bj{\"o}rnsson}, C.-I. 2013, \apj, 769, 65

\bibitem[\protect\citeauthoryear{{Bj{\"o}rnsson} \& {Fransson}}{{Bj{\"o}rnsson}
  \& {Fransson}}{2004}]{Bjornsson2004}
{Bj{\"o}rnsson}, C.-I.,  \& {Fransson}, C. 2004, \apj, 605, 823

\bibitem[\protect\citeauthoryear{{Bloom} et~al.}{{Bloom}
  et~al.}{2012}]{Bloom2012}
{Bloom}, J.~S., et~al. 2012, \pasp, 124, 1175

\bibitem[\protect\citeauthoryear{{Bramich}}{{Bramich}}{2008}]{Bramich2008}
{Bramich}, D.~M. 2008, \mnras, 386, L77

\bibitem[\protect\citeauthoryear{{Cappa}, {Goss}, \& {van der Hucht}}{{Cappa}
  et~al.}{2004}]{Cappa2004}
{Cappa}, C., {Goss}, W.~M.,  \& {van der Hucht}, K.~A. 2004, \aj, 127, 2885

\bibitem[\protect\citeauthoryear{{Cenko} et~al.}{{Cenko}
  et~al.}{2006}]{Cenko2006}
{Cenko}, S.~B., et~al. 2006, \pasp, 118, 1396

\bibitem[\protect\citeauthoryear{{Cenko} et~al.}{{Cenko}
  et~al.}{2013}]{Cenko2013}
{Cenko}, S.~B., et~al. 2013, \apj, 769, 130

\bibitem[\protect\citeauthoryear{{Chevalier}}{{Chevalier}}{1982}]{Chevalier198%
2}
{Chevalier}, R.~A. 1982, \apj, 259, 302

\bibitem[\protect\citeauthoryear{{Chevalier}}{{Chevalier}}{1996}]{Chevalier199%
6}
{Chevalier}, R.~A. 1996, in Astronomical Society of the Pacific Conference
  Series, Vol.~93, Radio Emission from the Stars and the Sun, ed. A.~R.
  {Taylor} \& J.~M. {Paredes}, 125

\bibitem[\protect\citeauthoryear{{Chevalier}}{{Chevalier}}{1998}]{Chevalier199%
8}
{Chevalier}, R.~A. 1998, \apj, 499, 810

\bibitem[\protect\citeauthoryear{{Chevalier} \& {Fransson}}{{Chevalier} \&
  {Fransson}}{2001}]{Chevalier2001}
{Chevalier}, R.~A.,  \& {Fransson}, C. 2001, ArXiv: astro-ph/0110060

\bibitem[\protect\citeauthoryear{{Chevalier} \& {Fransson}}{{Chevalier} \&
  {Fransson}}{2006}]{Chevalier2006}
{Chevalier}, R.~A.,  \& {Fransson}, C. 2006, ApJ, 651, 381

\bibitem[\protect\citeauthoryear{{Chevalier}, {Li}, \& {Fransson}}{{Chevalier}
  et~al.}{2004}]{Chevalier2004}
{Chevalier}, R.~A., {Li}, Z.-Y.,  \& {Fransson}, C. 2004, \apj, 606, 369

\bibitem[\protect\citeauthoryear{{Clocchiatti} et~al.}{{Clocchiatti}
  et~al.}{2011}]{Clocchiatti2011}
{Clocchiatti}, A., {Suntzeff}, N.~B., {Covarrubias}, R.,  \& {Candia}, P. 2011,
  \aj, 141, 163

\bibitem[\protect\citeauthoryear{{Clocchiatti} et~al.}{{Clocchiatti}
  et~al.}{1996}]{Clocchiatti1996}
{Clocchiatti}, A., {Wheeler}, J.~C., {Brotherton}, M.~S., {Cochran}, A.~L.,
  {Wills}, D., {Barker}, E.~S.,  \& {Turatto}, M. 1996, \apj, 462, 462

\bibitem[\protect\citeauthoryear{{Corsi} et~al.}{{Corsi}
  et~al.}{2012}]{Corsi2012}
{Corsi}, A., et~al. 2012, \apjl, 747, L5

\bibitem[\protect\citeauthoryear{{Draine}}{{Draine}}{1981}]{Draine1981}
{Draine}, B.~T. 1981, \apj, 245, 880

\bibitem[\protect\citeauthoryear{{Draine} \& {Lee}}{{Draine} \&
  {Lee}}{1984}]{Draine1984}
{Draine}, B.~T.,  \& {Lee}, H.~M. 1984, \apj, 285, 89

\bibitem[\protect\citeauthoryear{{Draine} \& {Salpeter}}{{Draine} \&
  {Salpeter}}{1979}]{Draine1979}
{Draine}, B.~T.,  \& {Salpeter}, E.~E. 1979, \apj, 231, 77

\bibitem[\protect\citeauthoryear{{Drout} et~al.}{{Drout}
  et~al.}{2011}]{Drout2011}
{Drout}, M.~R., et~al. 2011, \apj, 741, 97

\bibitem[\protect\citeauthoryear{{Dwek}}{{Dwek}}{1983}]{Dwek1983}
{Dwek}, E. 1983, \apj, 274, 175

\bibitem[\protect\citeauthoryear{{Dwek}}{{Dwek}}{1985}]{Dwek1985}
{Dwek}, E. 1985, \apj, 297, 719

\bibitem[\protect\citeauthoryear{{Ellison}, {Berezhko}, \& {Baring}}{{Ellison}
  et~al.}{2000}]{Ellison2000}
{Ellison}, D.~C., {Berezhko}, E.~G.,  \& {Baring}, M.~G. 2000, \apj, 540, 292

\bibitem[\protect\citeauthoryear{{Faber} et~al.}{{Faber} et~al.}{2003}]{DEIMOS}
{Faber}, S.~M., et~al. 2003, in Society of Photo-Optical Instrumentation
  Engineers (SPIE) Conference Series, Vol. 4841, Society of Photo-Optical
  Instrumentation Engineers (SPIE) Conference Series, ed. M.~{Iye} \& A.~F.~M.
  {Moorwood}, 1657

\bibitem[\protect\citeauthoryear{{Fazio} et~al.}{{Fazio}
  et~al.}{2004}]{Fazio2004}
{Fazio}, G.~G., et~al. 2004, \apjs, 154, 10

\bibitem[\protect\citeauthoryear{{Filippenko}}{{Filippenko}}{1997}]{Filippenko%
1997}
{Filippenko}, A.~V. 1997, \araa, 35, 309

\bibitem[\protect\citeauthoryear{{Filippenko} \& {Sargent}}{{Filippenko} \&
  {Sargent}}{1986}]{Filippenko1986}
{Filippenko}, A.~V.,  \& {Sargent}, W.~L.~W. 1986, \aj, 91, 691

\bibitem[\protect\citeauthoryear{{Foley} et~al.}{{Foley}
  et~al.}{2007}]{Foley2007}
{Foley}, R.~J., {Smith}, N., {Ganeshalingam}, M., {Li}, W., {Chornock}, R.,  \&
  {Filippenko}, A.~V. 2007, \apjl, 657, L105

\bibitem[\protect\citeauthoryear{{Fox} et~al.}{{Fox} et~al.}{2010}]{Fox2010}
{Fox}, O.~D., {Chevalier}, R.~A., {Dwek}, E., {Skrutskie}, M.~F., {Sugerman},
  B.~E.~K.,  \& {Leisenring}, J.~M. 2010, \apj, 725, 1768

\bibitem[\protect\citeauthoryear{{Fox} et~al.}{{Fox} et~al.}{2011}]{Fox2011}
{Fox}, O.~D., et~al. 2011, \apj, 741, 7

\bibitem[\protect\citeauthoryear{{Gal-Yam}, {Ofek}, \& {Shemmer}}{{Gal-Yam}
  et~al.}{2002}]{galyam2002}
{Gal-Yam}, A., {Ofek}, E.~O.,  \& {Shemmer}, O. 2002, \mnras, 332, L73

\bibitem[\protect\citeauthoryear{{Galama} et~al.}{{Galama}
  et~al.}{1998}]{Galama1998}
{Galama}, T.~J., et~al. 1998, \nat, 395, 670

\bibitem[\protect\citeauthoryear{{Gehrels} et~al.}{{Gehrels}
  et~al.}{2004}]{Gehrels2004}
{Gehrels}, N., et~al. 2004, \apj, 611, 1005

\bibitem[\protect\citeauthoryear{{Horesh} et~al.}{{Horesh}
  et~al.}{2012}]{Horesh2012}
{Horesh}, A., et~al. 2012, ArXiv e-prints 1209.1102

\bibitem[\protect\citeauthoryear{{Howell} et~al.}{{Howell}
  et~al.}{2005}]{Howell2005}
{Howell}, D.~A., et~al. 2005, ApJ, 634, 1190

\bibitem[\protect\citeauthoryear{{Immler} et~al.}{{Immler}
  et~al.}{2008}]{Immler2008}
{Immler}, S., et~al. 2008, \apjl, 674, L85

\bibitem[\protect\citeauthoryear{{Krauss} et~al.}{{Krauss}
  et~al.}{2012}]{Krauss2012}
{Krauss}, M.~I., et~al. 2012, \apjl, 750, L40

\bibitem[\protect\citeauthoryear{{Kulkarni} et~al.}{{Kulkarni}
  et~al.}{1998}]{Kulkarni1998}
{Kulkarni}, S.~R., et~al. 1998, \nat, 395, 663

\bibitem[\protect\citeauthoryear{{Laor} \& {Draine}}{{Laor} \&
  {Draine}}{1993}]{Laor1993}
{Laor}, A.,  \& {Draine}, B.~T. 1993, \apj, 402, 441

\bibitem[\protect\citeauthoryear{{Law} et~al.}{{Law} et~al.}{2009}]{Law2009}
{Law}, N.~M., et~al. 2009, \pasp, 121, 1395

\bibitem[\protect\citeauthoryear{{Li} \& {Song}}{{Li} \& {Song}}{2004}]{Li2004}
{Li}, Z.,  \& {Song}, L.~M. 2004, \apjl, 614, L17

\bibitem[\protect\citeauthoryear{{Li} \& {Chevalier}}{{Li} \&
  {Chevalier}}{1999}]{Li1999}
{Li}, Z.-Y.,  \& {Chevalier}, R.~A. 1999, \apj, 526, 716

\bibitem[\protect\citeauthoryear{{Maguire} et~al.}{{Maguire}
  et~al.}{2012}]{Kate2012}
{Maguire}, K., et~al. 2012, \mnras, 426, 2359

\bibitem[\protect\citeauthoryear{{Mauerhan} et~al.}{{Mauerhan}
  et~al.}{2013}]{Mauerhan2013}
{Mauerhan}, J.~C., et~al. 2013, \mnras, 430, 1801

\bibitem[\protect\citeauthoryear{{Mazzali} et~al.}{{Mazzali}
  et~al.}{2002}]{Mazzali2002}
{Mazzali}, P.~A., et~al. 2002, \apjl, 572, L61

\bibitem[\protect\citeauthoryear{{Mazzali} et~al.}{{Mazzali}
  et~al.}{2008}]{Mazzali2008}
{Mazzali}, P.~A., et~al. 2008, Science, 321, 1185

\bibitem[\protect\citeauthoryear{{McKenzie} \& {Schaefer}}{{McKenzie} \&
  {Schaefer}}{1999}]{McKenzie1999}
{McKenzie}, E.~H.,  \& {Schaefer}, B.~E. 1999, \pasp, 111, 964

\bibitem[\protect\citeauthoryear{{Moriya}, {Groh}, \& {Meynet}}{{Moriya}
  et~al.}{2013}]{Moriya2013}
{Moriya}, T.~J., {Groh}, J.~H.,  \& {Meynet}, G. 2013, ArXiv e-prints

\bibitem[\protect\citeauthoryear{{Nakano} et~al.}{{Nakano}
  et~al.}{2006}]{Nakano2006}
{Nakano}, S., {Itagaki}, K., {Puckett}, T.,  \& {Gorelli}, R. 2006, Central
  Bureau Electronic Telegrams, 666, 1

\bibitem[\protect\citeauthoryear{{Ofek} et~al.}{{Ofek}
  et~al.}{2013a}]{Ofek2013}
{Ofek}, E.~O., et~al. 2013a, \apj, 763, 42

\bibitem[\protect\citeauthoryear{{Ofek} et~al.}{{Ofek} et~al.}{2011}]{Eran2011}
{Ofek}, E.~O., {Frail}, D.~A., {Breslauer}, B., {Kulkarni}, S.~R., {Chandra},
  P., {Gal-Yam}, A., {Kasliwal}, M.~M.,  \& {Gehrels}, N. 2011, \apj, 740, 65

\bibitem[\protect\citeauthoryear{{Ofek} et~al.}{{Ofek} et~al.}{2012}]{Ofek2012}
{Ofek}, E.~O., et~al. 2012, \pasp, 124, 62

\bibitem[\protect\citeauthoryear{{Ofek} et~al.}{{Ofek}
  et~al.}{2013b}]{Ofek2013b}
{Ofek}, E.~O., {Lin}, L., {Kouveliotou}, C., {Younes}, G., {G{\"o}{\v
  g}{\"u}{\c s}}, E., {Kasliwal}, M.~M.,  \& {Cao}, Y. 2013b, \apj, 768, 47

\bibitem[\protect\citeauthoryear{{Ofek} et~al.}{{Ofek}
  et~al.}{2013c}]{Ofek2013a}
{Ofek}, E.~O., et~al. 2013c, \nat, 494, 65

\bibitem[\protect\citeauthoryear{{Oke} et~al.}{{Oke} et~al.}{1995}]{LRIS}
{Oke}, J.~B., et~al. 1995, PASP, 107, 375

\bibitem[\protect\citeauthoryear{{Panaitescu} \& {Kumar}}{{Panaitescu} \&
  {Kumar}}{2002}]{Panaitescu2002}
{Panaitescu}, A.,  \& {Kumar}, P. 2002, \apj, 571, 779

\bibitem[\protect\citeauthoryear{{Pastorello} et~al.}{{Pastorello}
  et~al.}{2007}]{Pastorello2007}
{Pastorello}, A., et~al. 2007, \nat, 447, 829

\bibitem[\protect\citeauthoryear{{Patat} et~al.}{{Patat}
  et~al.}{2001}]{Patat2001}
{Patat}, F., et~al. 2001, ApJ, 555, 900

\bibitem[\protect\citeauthoryear{{Perley} et~al.}{{Perley}
  et~al.}{2009}]{Perley2009}
{Perley}, R., et~al. 2009, IEEE Proceedings, 97, 1448

\bibitem[\protect\citeauthoryear{{Pian} et~al.}{{Pian} et~al.}{1999}]{Pian1999}
{Pian}, E., et~al. 1999, \aaps, 138, 463

\bibitem[\protect\citeauthoryear{{Poznanski}, {Prochaska}, \&
  {Bloom}}{{Poznanski} et~al.}{2012}]{Poznanski2012}
{Poznanski}, D., {Prochaska}, J.~X.,  \& {Bloom}, J.~S. 2012, MNRAS, 426, 1465

\bibitem[\protect\citeauthoryear{{Prieto} et~al.}{{Prieto}
  et~al.}{2013}]{Prieto2013}
{Prieto}, J.~L., {Brimacombe}, J., {Drake}, A.~J.,  \& {Howerton}, S. 2013,
  \apjl, 763, L27

\bibitem[\protect\citeauthoryear{{Rau} et~al.}{{Rau} et~al.}{2009}]{Rau2009}
{Rau}, A., et~al. 2009, \pasp, 121, 1334

\bibitem[\protect\citeauthoryear{{Readhead}}{{Readhead}}{1994}]{Readhead1994}
{Readhead}, A.~C.~S. 1994, \apj, 426, 51

\bibitem[\protect\citeauthoryear{{Richmond} et~al.}{{Richmond}
  et~al.}{1996}]{Richmond1996}
{Richmond}, M.~W., et~al. 1996, \aj, 111, 327

\bibitem[\protect\citeauthoryear{{Salas} et~al.}{{Salas}
  et~al.}{2013}]{Salas2012}
{Salas}, P., {Bauer}, F.~E., {Stockdale}, C.,  \& {Prieto}, J.~L. 2013, \mnras,
  428, 1207

\bibitem[\protect\citeauthoryear{{Schlafly} \& {Finkbeiner}}{{Schlafly} \&
  {Finkbeiner}}{2011}]{Schlafly2011}
{Schlafly}, E.~F.,  \& {Finkbeiner}, D.~P. 2011, \apj, 737, 103

\bibitem[\protect\citeauthoryear{{Smith}, {Foley}, \& {Filippenko}}{{Smith}
  et~al.}{2008}]{Smith2008}
{Smith}, N., {Foley}, R.~J.,  \& {Filippenko}, A.~V. 2008, \apj, 680, 568

\bibitem[\protect\citeauthoryear{{Soderberg} et~al.}{{Soderberg}
  et~al.}{2008}]{Soderberg2008}
{Soderberg}, A.~M., et~al. 2008, \nat, 453, 469

\bibitem[\protect\citeauthoryear{{Soderberg} et~al.}{{Soderberg}
  et~al.}{2010}]{Soderberg2010}
{Soderberg}, A.~M., et~al. 2010, \nat, 463, 513

\bibitem[\protect\citeauthoryear{{Soderberg} et~al.}{{Soderberg}
  et~al.}{2006a}]{Soderberg2003bg}
{Soderberg}, A.~M., {Chevalier}, R.~A., {Kulkarni}, S.~R.,  \& {Frail}, D.~A.
  2006a, \apj, 651, 1005

\bibitem[\protect\citeauthoryear{{Soderberg} et~al.}{{Soderberg}
  et~al.}{2005}]{Soderberg2003L}
{Soderberg}, A.~M., {Kulkarni}, S.~R., {Berger}, E., {Chevalier}, R.~A.,
  {Frail}, D.~A., {Fox}, D.~B.,  \& {Walker}, R.~C. 2005, \apj, 621, 908

\bibitem[\protect\citeauthoryear{{Soderberg} et~al.}{{Soderberg}
  et~al.}{2006b}]{Soderberg2006}
{Soderberg}, A.~M., {Nakar}, E., {Berger}, E.,  \& {Kulkarni}, S.~R. 2006b,
  \apj, 638, 930

\bibitem[\protect\citeauthoryear{{Steele} et~al.}{{Steele}
  et~al.}{2004}]{Steele2004}
{Steele}, I.~A., et~al. 2004, in Society of Photo-Optical Instrumentation
  Engineers (SPIE) Conference Series, Vol. 5489, Society of Photo-Optical
  Instrumentation Engineers (SPIE) Conference Series, ed. J.~M. {Oschmann},
  Jr., 679

\bibitem[\protect\citeauthoryear{{Sutaria} et~al.}{{Sutaria}
  et~al.}{2003}]{Sutaria2003}
{Sutaria}, F.~K., {Chandra}, P., {Bhatnagar}, S.,  \& {Ray}, A. 2003, \aap,
  397, 1011

\bibitem[\protect\citeauthoryear{{van Eerten} \& {MacFadyen}}{{van Eerten} \&
  {MacFadyen}}{2011}]{van2011}
{van Eerten}, H.~J.,  \& {MacFadyen}, A.~I. 2011, ApJL, 733, L37

\bibitem[\protect\citeauthoryear{{Walker}}{{Walker}}{1998}]{Walker1998}
{Walker}, M.~A. 1998, \mnras, 294, 307

\bibitem[\protect\citeauthoryear{{Walker}}{{Walker}}{2001}]{Walker2001}
{Walker}, M.~A. 2001, \mnras, 321, 176

\bibitem[\protect\citeauthoryear{{Wang}, {Huang}, \& {Kong}}{{Wang}
  et~al.}{2009}]{Wang2009}
{Wang}, X., {Huang}, Y.~F.,  \& {Kong}, S.~W. 2009, \aap, 505, 1213

\bibitem[\protect\citeauthoryear{{Weiler}, {Panagia}, \& {Sramek}}{{Weiler}
  et~al.}{1990}]{Weiler1990}
{Weiler}, K.~W., {Panagia}, N.,  \& {Sramek}, R.~A. 1990, \apj, 364, 611

\bibitem[\protect\citeauthoryear{{Weiler} et~al.}{{Weiler}
  et~al.}{1986}]{Weiler1986}
{Weiler}, K.~W., {Sramek}, R.~A., {Panagia}, N., {van der Hulst}, J.~M.,  \&
  {Salvati}, M. 1986, \apj, 301, 790

\bibitem[\protect\citeauthoryear{{Weisskopf} et~al.}{{Weisskopf}
  et~al.}{2002}]{Weiss2002}
{Weisskopf}, M.~C., {Brinkman}, B., {Canizares}, C., {Garmire}, G., {Murray},
  S.,  \& {Van Speybroeck}, L.~P. 2002, \pasp, 114, 1

\bibitem[\protect\citeauthoryear{{Wellons}, {Soderberg}, \&
  {Chevalier}}{{Wellons} et~al.}{2012}]{Wellons2012}
{Wellons}, S., {Soderberg}, A.~M.,  \& {Chevalier}, R.~A. 2012, \apj, 752, 17

\bibitem[\protect\citeauthoryear{{Woosley} \& {Bloom}}{{Woosley} \&
  {Bloom}}{2006}]{Bloom2006}
{Woosley}, S.~E.,  \& {Bloom}, J.~S. 2006, \araa, 44, 507

\bibitem[\protect\citeauthoryear{{Yaron} \& {Gal-Yam}}{{Yaron} \&
  {Gal-Yam}}{2012}]{Yaron2012}
{Yaron}, O.,  \& {Gal-Yam}, A. 2012, \pasp, 124, 668

\bibitem[\protect\citeauthoryear{{York} et~al.}{{York} et~al.}{2000}]{York2000}
{York}, D.~G., et~al. 2000, The Astronomical Journal, 120, 1579

\bibitem[\protect\citeauthoryear{{Yost} et~al.}{{Yost} et~al.}{2003}]{Yost2003}
{Yost}, S.~A., {Harrison}, F.~A., {Sari}, R.,  \& {Frail}, D.~A. 2003, \apj,
  597, 459

\end{thebibliography}
\newpage

\begin{center}
\begin{longtable}{lcccc}
\caption{Optical observations of PTF11qcj. \label{opt}}
\\
\hline
\hline
Date & Telescope & Band & Exposure Time & Magnitude \\
(MJD) & & & (s) & [AB]\\ 
\hline
\endhead
\hline
 55857.543 & P48 & $g$ & 60 &$17.579\pm0.040$\\
 55859.538 & P48 & $g$ & 60 & $17.726\pm0.022$\\
 55862.530 & P48 & $g$ & 60 &$18.046\pm0.018$\\
 55864.524 & P48 & $g$ & 60 &$18.189\pm0.024$\\
 55865.520 & P48 & $g$ & 60 &$18.283\pm0.019$\\
...\footnote{This Table is published in its entirety in the electronic version of this paper; a portion is shown here for guidance regarding its form and content.}&...&...&...&...\\
\hline
\end{longtable}
\end{center}
\begin{center}
\begin{longtable}{lcccc}
\caption{Radio observations of PTF11qcj.\label{radioTab}}
\\
\hline
\hline
MJD & Observatory & Central freq. & Flux Density \\
          &      & (GHz)    & (mJy/beam)  \\
\hline
\endhead
55880.538 & VLA - D & 6 & $0.577\pm0.041$\\
55880.538 & VLA - D & 5 & $0.421\pm0.026$\\
\hline
55884.803 & CARMA & 93 & $3.96\pm0.88$\\
\hline
55887.508 & VLA - D &  7.4  & $1.556\pm0.079$  \\
55887.508 & VLA - D &  5    & $0.619\pm0.034$ \\
\hline
55891.640 & CARMA & 93 & $3.62\pm0.75$\\
...\footnote{This Table is published in its entirety in the electronic version of this paper; a portion is shown here for guidance regarding its form and content.}&...&...&...\\
\hline
\hline
\end{longtable}
\end{center}
\begin{center}
\begin{longtable}{lccccc}
\caption{X-ray observations of PTF11qcj. Count rates have been converted into fluxes assuming a spectral index of $\Gamma\approx 1.7$ and a Galactic column density $N_H\approx 10^{20}$\,cm$^{-2}$.\label{X}}
\\
\hline
\hline
Date & Instrument & Band & Exposure Time & Count Rate & Flux (0.3-8\,keV, unabs)\\
(MJD) & & (keV) & (ks) & (s$^{-1}$) & (erg\,cm$^{-2}$\,s$^{-1}$)\\ 
\hline
\endhead
55883.00 & \textit{Swift}-XRT &  0.2-10 & 4.352 & $<6.9\times10^{-4}$ & $<2.4\times10^{-14}$\\
55916.71 & \textit{Swift}-XRT & 0.2-10 & 4.937 & $<9.6\times10^{-4}$ & $<3.4\times10^{-14}$\\
55923.59 & \textit{Swift}-XRT & 0.2-10 & 1.216 & $<2.5\times10^{-3}$ & $<8.8\times10^{-14}$\\
55930.02 & \textit{Swift}-XRT & 0.2-10 & 1.538 & $<3.1\times10^{-3}$ & $<1.1\times10^{-13}$ \\
55937.63 & \textit{Swift}-XRT & 0.2-10 & 4.750 & $<6.3\times10^{-4}$ & $<2.2\times10^{-14}$ \\
55949.07 & \textit{Swift}-XRT & 0.2-10 & 9.231 & $<6.8\times10^{-4}$ & $<2.4\times10^{-14}$ \\
55956.02 & \textit{Swift}-XRT & 0.2-10 & 2.951 & $<1.0\times10^{-3}$ & $<3.5\times10^{-14}$\\
55957.09 & \textit{Swift}-XRT & 0.2-10 & 2.402 & $<2.0\times10^{-3}$ & $<7.0\times10^{-14}$ \\
55930 (mean) & \textit{Swift}-XRT & 0.2-10 & 31.380 & $< 4.1\times10^{-4}$ & $<1.4\times10^{-14}$\\ 
55939.05 & \textit{Chandra}-ACIS & 0.3-8  & 9.836  & $(8.8\pm3.0)\times10^{-4}$ & $(7.6\pm2.6)\times10^{-15}$\\
55983.37 & \textit{Chandra}-ACIS & 0.3-8  & 9.931 & $(5.7\pm2.5)\times10^{-4}$ & $(4.9\pm2.1)\times10^{-15}$\\
56028.36 & \textit{Chandra}-ACIS & 0.3-8  & 19.930 & $(3.3\pm1.3)\times10^{-4}$ & $(3.1\pm1.2)\times10^{-15}$\\
\hline
\end{longtable}
\end{center}

\begin{center}
\begin{longtable}{lcccc}
\caption{IR observations of PTF11qcj \label{spitzertab}}
\\
\hline
\hline
MJD  & Observatory& Wavelength ($\mu$m) &  Flux density ($\mu$Jy)\\
\hline
\endhead
56014.063 & P200             & 2.159 & $53\pm4$\\
56014.747 & \textit{Spitzer} & 3.550 & $195.0\pm1.7$\\
56014.747 & \textit{Spitzer} & 4.493 & $158.3\pm1.2$\\
56103.643 & \textit{Spitzer} & 3.550 & $177.9\pm1.6$\\
56103.643 & \textit{Spitzer} & 4.493 & $144.3\pm1.1$\\
\hline
\end{longtable}
\end{center}
\end{document}